\documentclass[aps,prb,twocolumn,showpacs,groupedaddress, amsmath]{revtex4}

\usepackage{graphicx}
\usepackage{amssymb}

\begin{document}


\title{Memory and aging effect in hierarchical spin orderings of stage-2 CoCl$_{2}$ graphite intercalation compound}


\author{Masatsugu Suzuki}
\email[]{suzuki@binghamton.edu}
\affiliation{Department of Physics, State University of New York at Binghamton, Binghamton, New York 13902-6000}

\author{Itsuko S. Suzuki}
\email[]{itsuko@binghamton.edu}
\affiliation{Department of Physics, State University of New York at Binghamton, Binghamton, New York 13902-6000}

\author{Motohiro Matsuura}
\affiliation{Department of Management and Information Science, Fukui University of Technology, Fukui, Fukui 910-8505, JAPAN}


\date{\today}

\begin{abstract}
Stage-2 CoCl$_{2}$ graphite intercalation compound undergoes two magnetic phase transitions at $T_{cl}$ (= 7.0 K) and $T_{cu}$ (= 8.9 K). The aging dynamics of this compound is studied near $T_{cl}$ and $T_{cu}$. The intermediate state between $T_{cl}$ and $T_{cu}$ is characterized by a spin glass phase extending over ferromagnetic islands. A genuine thermoremnant magnetization (TRM) measurement indicates that the memory of the specific spin configurations imprinted at temperatures between $T_{cl}$ and $T_{cu}$ during the field-cooled (FC) aging protocol can be recalled when the system is re-heated at a constant heating rate. The zero-field cooled (ZFC) and TRM magnetization is examined in a series of heating and reheating process. The magnetization shows both characteristic memory and rejuvenation effects. The time $(t)$ dependence of the relaxation rate $S_{ZFC}(t)=(1/H)$d$M_{ZFC}(t)$/d$\ln t$ after the ZFC aging protocol with a wait time $t_{w}$, exhibits two peaks at characteristic times $t_{cr1}$ and $t_{cr2}$ between $T_{cl}$ and $T_{cu}$. An aging process is revealed as the strong $t_{w}$ dependence of $t_{cr2}$. The observed aging and memory effect is discussed in terms of the droplet model.
\end{abstract}

\pacs{75.40.Gb, 75.50.Lk, 75.30.Kz, 75.30.GW}

\maketitle



\section{\label{intro}INTRODUCTION}
Magnetic phase transitions of stage-2 CoCl$_{2}$ graphite intercalation compound (GIC) have been extensively studied.\cite{Enoki2003,Murakami1983,Matsuura1987a,Wiesler1987,Matsuura1987b,Murakami1988,Matsuura1990,Matsuura1992,Miyoshi1996,Suzuki1998,Suzuki2002} This compound magnetically behaves like a quasi two-dimensional (2D) XY-like ferromagnet with a very weak antiferromagnetic interplanar interaction. The intercalate layers are formed of small islands whose average diameters are on the order of 450 $\AA$. The peripheral chlorine ions at the island boundary provide acceptor sites for charges transferred from the graphite layer to the intercalate layer. This compound undergoes magnetic phase transitions at $T_{cu}$ (= 8.9 K) and $T_{cl}$ (= 7.0 K). The growth of the in-plane spin correlation length $\xi_{a}$ is limited by the existence of the islands, making the effective interplanar exchange interaction finite and suppressing the 3D spin ordering to a lower temperature than $T_{cu}$. At $T_{cu}$ a 2D ferromagnetic spin order develops inside each island. Between $T_{cl}$ and $T_{cu}$ the 2D ferromagnetic long rang order (LRO) is established. The in-plane spin correlation length grows to the order of the island size at $T_{cl}$. Below $T_{cl}$ the system is in a 3D antiferromagnetic phase with the 2D ferromagnetic layers being stacked antiferromagnetically along the \textit{c} axis.

However, it seems that such a simple picture for the ordering process below $T_{cu}$, is not appropriate for the peculiar phenomena observed so far in stage-2 CoCl$_{2}$ GIC,\cite{Matsuura1987a,Matsuura1987b,Matsuura1990,Matsuura1992,Miyoshi1996} which is rather characteristic of the spin glass (SG) phase. Further it may be also inconsistent with the following experimental results. (i) The absorption $\chi^{\prime\prime}$ shows peaks at $T_{cl}$ and $T_{cu}$. The peak at $T_{cl}$ shifts to the high-$T$ side with increasing $f$, while the peak at $T_{cu}$ remains unshifted.\cite{Suzuki1998} (ii) The spin correlation length along the $c$ axis, $\xi_{c}$, grows rapidly below $T_{cu}$ but quickly saturates to a constant value of 22 $\AA$, or less than two magnetic layers (the $c$-axis repeat distance $d=12.70 \AA$).\cite{Wiesler1987} Below $T_{cl}$ a 3D antiferromagnetic long range order (LRO) is established mainly through effective interplanar interactions including interisland interactions between islands in adjacent intercalate layers.

Under such a circumstance, we performed a series of experiments on the dynamical aspect of the ordering process. Most of them are used to study the aging dynamics (aging, memory and rejuvenation) of SG's. In this paper we study the aging dynamics of stage-2 CoCl$_{2}$ GIC near $T_{cu}$ and $T_{cl}$. Our experiments include (i) genuine thermoremnant magnetization (TRM), (ii) relaxation rate of the ZFC susceptibility, (iii) zero-field cooled (ZFC) magnetization and TRM magnetization in a series of heating and cooling process which is the same procedure used by Matsuura et al.\cite{Matsuura1987b} for stage-2 CoCl$_{2}$ GIC, and (iv) field-cooled magnetization in a FC cooling protocol (with an intermittent stop for a wait time in the absence of a magnetic field) which is the same protocol used by Sun et al.\cite{Sun2003} for permalloy (Ni$_{81}$Fe$_{19}$) nanoparticles.

We show that the intermediate state is a SG ordered phase extending over ferromagnetic islands. The magnetization shows both characteristic memory and rejuvenation effects. The time ($t$) dependence of the relaxation rate $S_{ZFC}(t)=(1/H)$d$M_{ZFC}(t)$/d$\ln t$ after the ZFC aging protocol with a wait time $t_{w}$, exhibits two peaks at characteristic times $t_{cr1}$ and $t_{cr2}$ between $T_{cl}$ and $T_{cu}$. An aging process is revealed as the strong $t_{w}$ dependence of $t_{cr2}$. The observed aging and memory effect will be discussed in terms of the droplet model.\cite{Fisher1986,Bray1987,Fisher1988} 

\section{\label{exp}EXPERIMENTAL PROCEDURE}
A stage-2 CoCl$_{2}$ GIC sample was prepared by intercalation of pristine CoCl$_{2}$ into a single crystal of kish graphite in a Cl$_{2}$ gas atmosphere at 740 Torr for three weeks at 540 $^{\circ}$C. The sample used in the present experiment is one used in the previous experiments.\cite{Suzuki1998,Suzuki2002} The DC magnetization and AC magnetic susceptibility were measured using a SQUID magnetometer (Quantum Design, MPMS XL-5) with an ultra low field capability option. Before the measurement, a remnant magnetic field was reduced to zero field (exactly less than 3 mOe) at 298 K. We measured the DC magnetization as a function of temperature after various kinds of cooling protocol. The detail of the cooling protocol for each experiment will be presented in Sec.~\ref{result}. We also measured the time dependence of the ZFC magnetization at various wait times.

\section{\label{result}RESULT}
\subsection{\label{resultA}$T$ dependence of $M_{ZFC}$, $M_{FC}$, $M_{TRM}$, $M_{IRM}$ and $\Delta M$ ($= M_{FC}-M_{ZFC}$)}
In the previous paper,\cite{Suzuki1998} we have studied the dynamic aspect of in-plane spin ordering in stage-2 CoCl$_{2}$ GIC from both the dispersion $\chi^{\prime}$ and absorption $\chi^{\prime\prime}$ at $f = 0.1$ Hz acquired using an AC magnetic susceptibility. The absorption $\chi^{\prime\prime}$ shows three peaks at $T = T_{cu}$ (= 8.9 K), $T_{p1}$ (= 8.4 K), and $T_{cl}$ (= 7.0 K), while $\chi^{\prime}$ has a single peak at $T_{p1}$. These results indicate that this compound undergoes two magnetic phase transitions at $T_{cl}$ and $T_{cu}$.

In order to see how these successive magnetic phase transitions are observed in the DC magnetic susceptibility, we measured the temperature ($T$) dependence of the magnetization $M_{ZFC}$, $M_{FC}$, and $M_{TRM}$ in the case of $H = 1$ and 0.15 Oe. (a) \textit{The zero-field cooled magnetization} ($M_{ZFC}$) \textit{measurement}. The system was annealed at 50 K for 1200 sec in the absence of $H$. The system was cooled from 50 to 1.9 K at $H$ = 0 (ZFC aging protocol). After the system was aged at 1.9 K for $t_{w}$ = 100 sec at $H$ = 0, the magnetic field is applied at $H$ (= 1 and 0.15 Oe). Subsequently $M_{ZFC}$ was measured with increasing $T$ from 1.9 to 12 K at the rate of 0.025 K/minute. (b) \textit{The field cooled magnetization} ($M_{FC}$) \textit{measurement}. The system was annealed at 50 K for 1200 sec in the presence of $H$ (= 1 and 0.15 Oe). Then the system was cooled from 50 to 12 K in the presence of $H$ (FC aging protocol). The magnetization $M_{FC}$ was measured with decreasing $T$ from 12 to 1.9 K. (c) \textit{The thermoremnant magnetization} ($M_{TRM}$) \textit{measurement}. The system was cooled from 50 to 1.9 K in the presence of $H=H_{c}$ (= 1 and 0.15 Oe) through the FC aging protocol. After the system was aged at 1.9 K for $t_{w}=100$ sec, the field was cut off ($H=0$). Then the magnetization $M_{TRM}$ was measured with increasing $T$ from 1.9 to 12 K.

\begin{figure}
\includegraphics[width=7.0cm]{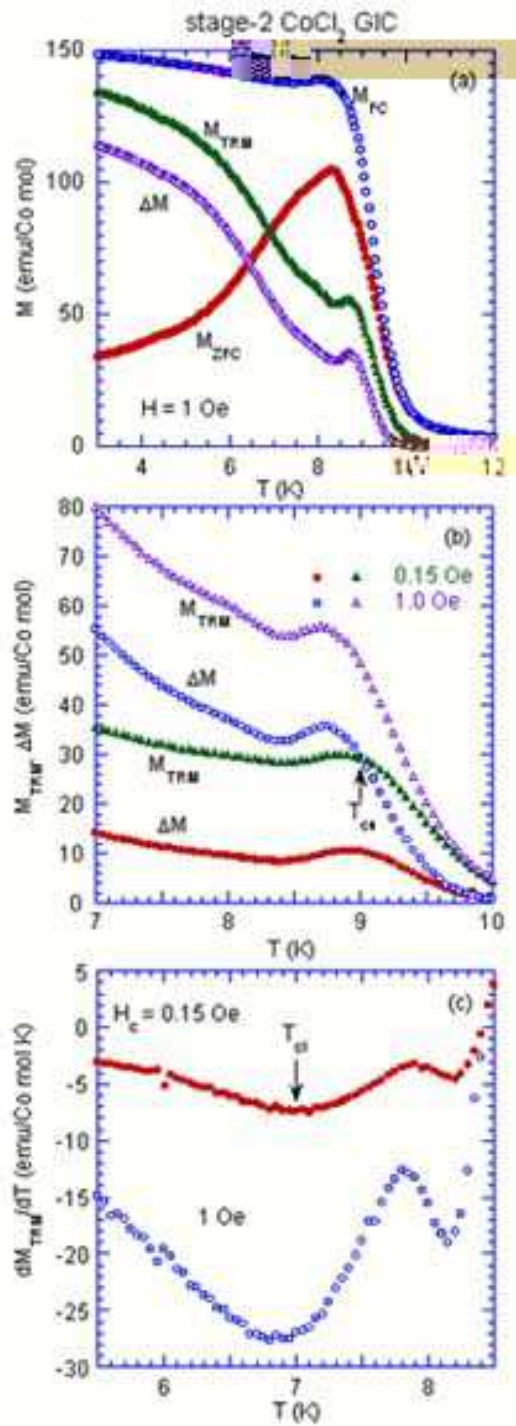}
\caption{\label{fig01}(Color online) (a) $T$ dependence of $M_{ZFC}$, $M_{FC}$, $M_{TRM}$ and $\Delta M$ (= $M_{FC}-M_{ZFC}$) for stage-2 CoCl$_{2}$ GIC. $H$ = 1 Oe. (b) $T$ dependence of $M_{TRM}$ at $H_{c}$ = 1 and 0.15 Oe and $\Delta M$ at $H$ = 1 and 0.15 Oe. $T_{cu}$ = 8.9 K (denoted by arrow). (c) $T$ dependence of d$M_{TRM}$/d$T$ at $H_{c}$= 1 and 0.15 Oe. $T_{cl}$ = 7.0 K (denoted by arrow).}
\end{figure}

Figure \ref{fig01}(a) shows the $T$ dependence of $M_{ZFC}$, $M_{FC}$, $M_{TRM}$, and $\Delta M=M_{FC}-M_{ZFC}$ for $H$ = 1 Oe. The magnetization $M_{ZFC}$ exhibits a peak at 8.3 K close to $T_{p1}$, which remains unchanged with decreasing $H$ from 1 to 0.15 Oe.  The deviation of $M_{ZFC}$ from $M_{FC}$ starts to occur below 10.5 K, due to the irreversibility effect of magnetization. Figure \ref{fig01}(b) shows the $T$ dependence of $M_{TRM}$ for $H_{c}$ = 0.15 and 1 Oe, and $\Delta M$ for $H$ = 0.15 and 1 Oe. The magnetization $M_{TRM}$ at $H_{c}$ = 1 Oe exhibits a local maximum at 8.70 K. The local-maximum temperature increases with decreasing $H_{c}$ and reaches 8.85 K at $H_{c}$ = 0.15, which is very close to $T_{cu}$. Although $\Delta M$ at $H=H_{0}$ (= 0.15 and 1 Oe) is smaller than $M_{TRM}$ at $H_{c}=H_{0}$, below 10 K, the $T$ dependence of $\Delta M$ at $H= H_{0}$ is similar to that of $M_{TRM}$ at $H=H_{0}$. Figure \ref{fig01}(c) shows the $T$ dependence of d$M_{TRM}$/d$T$ at $H_{c}=1$ and 0.15 Oe. The derivative d$M_{TRM}$/d$T$ exhibits a negative local minimum at $T=6.9$ K for $H_{c}=1$ Oe and at $T=T_{cu}$ (= 7.0 K) for $H_{c}=0.15$ Oe.

\subsection{\label{resultB}Genuine TRM measurement}

\begin{figure*}
\includegraphics[width=12.0cm]{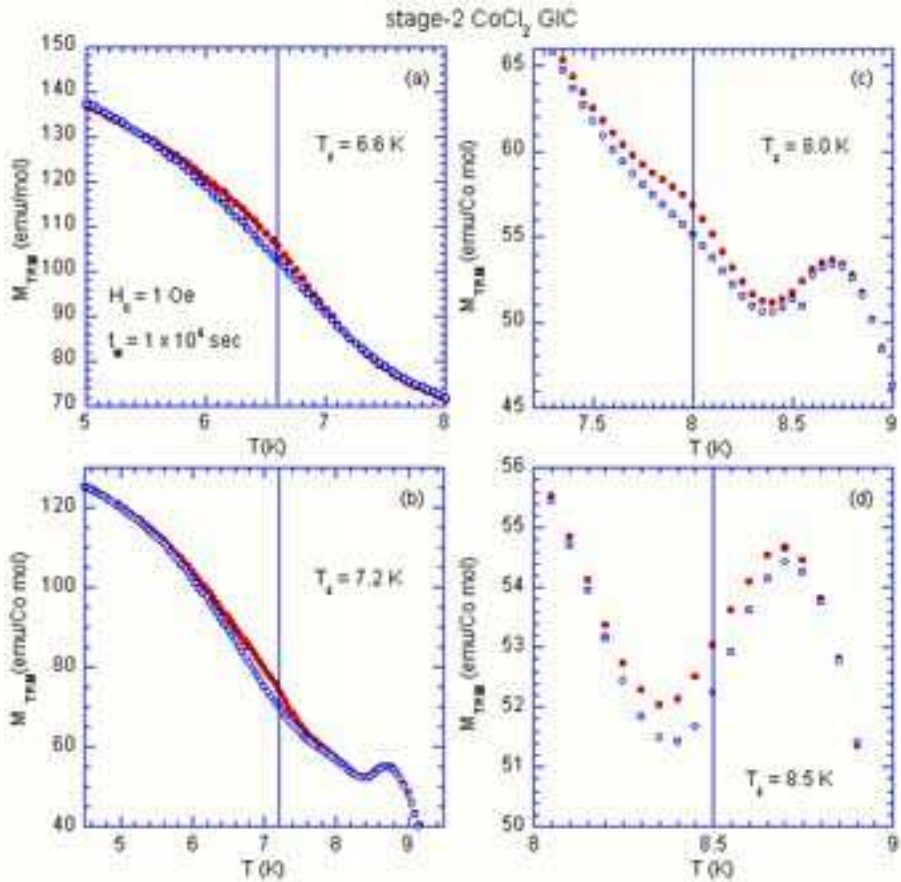}
\caption{\label{fig02}(Color online) $T$ dependence of $M_{TRM}(T;T_{s},t_{s})$ (closed circles) and of $M_{TRM}^{ref}(T)$ (open circles). $M_{TRM}(T;T_{s},t_{s})$ is measured with increasing $T$ at $H$ = 0 after the FC cooling protocol at $H_{c}$ = 1 Oe with a stop-wait at $T_{s}$ for $t_{s} = 1.0 \times 10^{4}$ sec. $M_{TRM}^{ref}$ is measured with increasing $T$ after the FC cooling protocol at $H_{c}$ = 1 Oe without such a stop-wait protocol. (a) $T_{s}$ = 6.6 K, (b) 7.2 K, (c) 8.0 K, and (d) 8.5 K.}
\end{figure*}

\begin{figure}
\includegraphics[width=6.5cm]{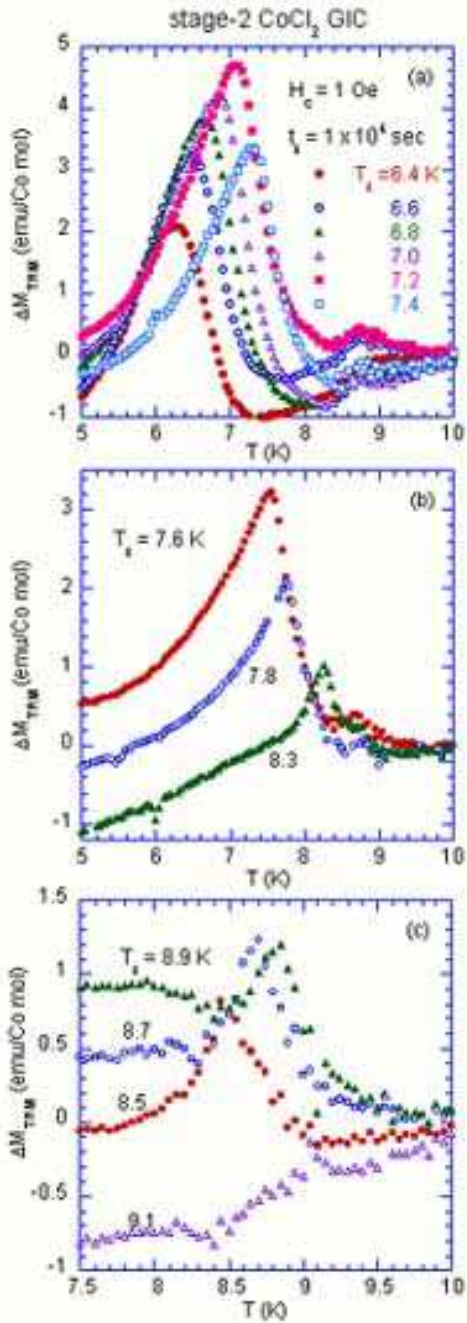}
\caption{\label{fig03}(Color online) $T$ dependence of $\Delta M_{TRM}(T;T_{s},t_{s})$ [$=M_{TRM}(T;T_{s},t_{s})-M_{TRM}^{ref}(T)$]. $t_{s} = 1.0 \times 10^{4}$ sec. $6.4\leq T_{s}\leq 9.1$ K. $H_{c}$ = 1 Oe. (a) $6.6\leq T\leq 7.4$ K. (b) $7.6\leq T \leq 8.3$ K. (c) $8.5\leq T \leq 9.1$ K.}
\end{figure}

\begin{figure}
\includegraphics[width=7.0cm]{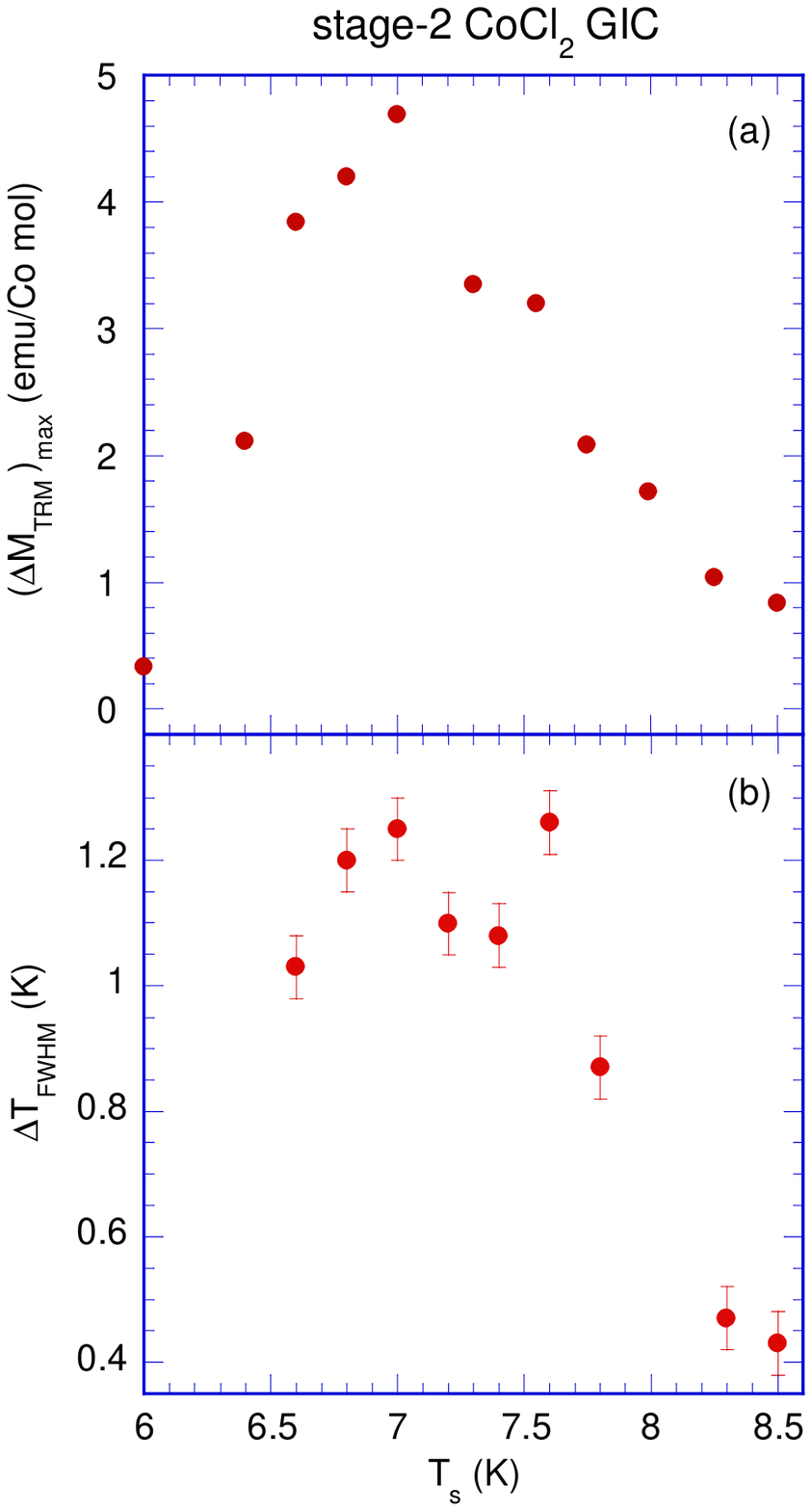}
\caption{\label{fig04}(Color online) (a) $T_{s}$ dependence of the peak height of $\Delta M_{TRM}(T;T_{s},t_{s})$ denoted by $(\Delta M_{TRM})_{max}$. The curve of $\Delta M_{TRM}$ vs $T$ shows a peak at $T_{s}$. (b) $T_{s}$ dependence of the full-width at half-maximum (= $\Delta T_{FWHM}$) for the peak of the curve of $\Delta M_{TRM}$ vs $T$ located at $T=T_{s}$.}
\end{figure}

We present a memory effect observed in the so-called genuine TRM magnetization measurement of stage-2 CoCl$_{2}$ GIC. Similar behavior is observed in SG's.\cite{Mathieu2001a,Mathieu2001b,Sahoo2003a,Sahoo2003b} Our system was cooled from 50 K to an intermittent stop temperature $T_{s}$ ($6\leq T_{s}\leq 9.1$ K) in the presence of $H=H_{c}$ (= 1 Oe) (stop-wait process). The system was aged at $T_{s}$ for a wait time $t_{s}$ ($=1.0\times 10^{4}$ sec) at $H=H_{c}$, and subsequently it was cooled again down to 3.0 K. After the magnetic field was switched off at 3 K. the TRM magnetization was measured with increasing $T$ from 3.0 to 12.0 K at $H$ = 0. This value of the TRM magnetization is compared with that of the TRM magnetization which was measured with increasing $T$ after the FC cooling protocol without any stop-wait process [$M_{TRM}^{ref}(T)$ as the reference curve]. Here we define the difference $\Delta M_{TRM}(T;T_{s},t_{s})$ as
\begin{equation} 
\Delta M_{TRM}(T,T_{s},t_{s})=M_{TRM}(T;T_{s},t_{s})-M_{TRM}^{ref}(T).
\label{eq01} 
\end{equation} 
Figure \ref{fig02} shows the $T$ dependence of $M_{TRM}(T)(T;T_{s},t_{s})$ (stop-wait curve) and $M_{TRM}^{ref}(T)$ (reference curve) at $t_{s}= 1.0 \times 10^{4}$ sec for typical stop temperatures ($T_{s}$ = 6.6, 7.2, 8.0, and 8.5 K). The reference curve and the stop-wait curves coalesce at low temperatures and only start to deviate as $T_{s}$ is approached from the low $T$ side. The stop-wait curve lie significantly above the reference curve in the vicinity of $T_{s}$. Figure \ref{fig03} shows the $T$ dependence of $\Delta M_{TRM}(T;T_{s},t_{s})$ at various $T_{s}$. The difference $\Delta M_{TRM}(T;T_{s},t_{s})$ shows either a symmetric broad peak centered at $T=T_{s}$ for $T_{s}<7.4$ K or an asymmetric cusp centered at $T=T_{s}$ for $7.6$ K $<T<T_{cu}$. The peak of $\Delta M_{TRM}(T;T_{s},t_{s}$) is not observed for $T_{s}\leq 6$ K. The result indicates that the spin configuration imprinted at $T_{s}$ is recovered on reheating. In this sense, the system sustains a memory of an equilibrium state reached after a stop-wait process at $T_{s}$. Such phenomena are commonly observed in various kinds of SG's. Figure \ref{fig04}(a) shows the $T_{s}$ dependence of the peak height of the curve $\Delta M_{TRM}(T;T_{s},t_{s})$ vs $T$ located at $T_{s}$, [denoted by $(\Delta M_{TRM})_{max}$]. The peak height exhibits a sharp peak at $T_{s}=T_{cl}$. It decreases with further increasing $T_{s}$ and tends to zero around $T_{cu}$. As will be discussed in Sec.~\ref{disC}, the magnetization $(\Delta M_{TRM})_{max}$ is roughly proportional to $\xi_{a}^{2}$ between $T_{cl}$ and $T_{cu}$. The increase of $(\Delta M_{TRM})_{max}$ with decreasing $T$ between $T_{cl}$ and $T_{cu}$ suggests that $\xi_{a}$ increases with decreasing $T$ from $T_{cu}$ to $T_{cl}$. Figure \ref{fig04}(b) shows the $T_{s}$ dependence of the full-width at half-maximum [$= (\Delta T)_{FWHM}$] for the peak of the curve $\Delta M_{TRM}$ vs $T$ located at $T=T_{s}$. The full-width at half-maximum $(\Delta T)_{FWHM}$ exhibits a peak around $T_{s}=T_{cl}$. It is on the order of 1.2 K at $T_{s}=T_{cl}$ and 0.4 K around $T_{s}=T_{cu}$. The $T_{s}$ dependence of ($\Delta T)_{FWHM}$ will be discussed in terms of the overlap length of the droplet model\cite{Fisher1986,Bray1987,Fisher1988} for SG's in Sec.~\ref{disD}.

\subsection{\label{resultC}Relaxation rate $S_{ZFC}(t)$ }

\begin{figure}
\includegraphics[width=7.0cm]{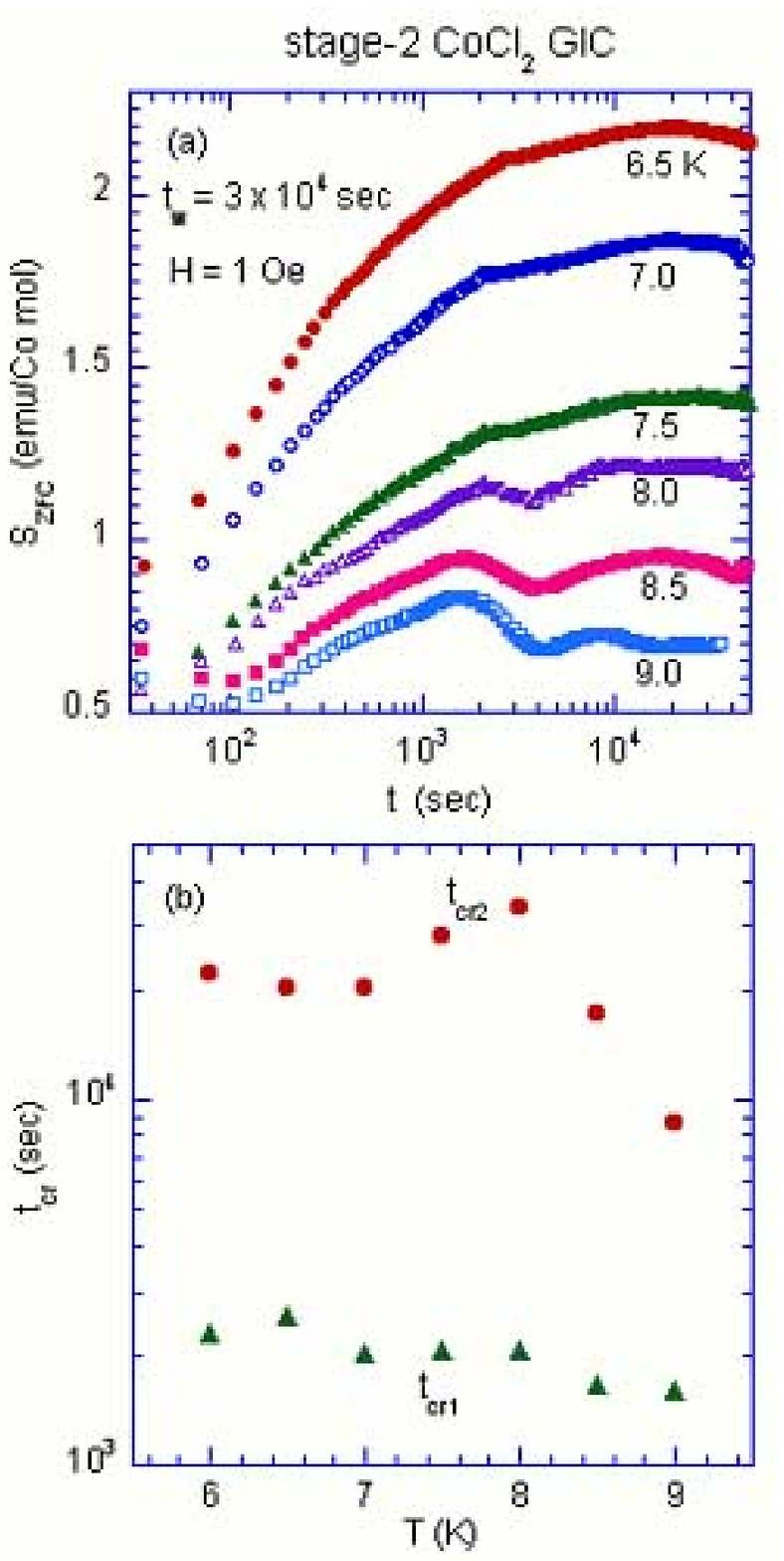}
\caption{\label{fig05}(Color online) (a) $t$ dependence of the relaxation rate $S_{ZFC}(t)$ [$=(1/H)$d$M_{ZFC}(t)$/d$\ln t$]. $M_{ZFC}(t)$ was measured at a fixed $T$ as a function of $t$ after the ZFC cooling protocol consisting of the cooling from 50 K to $T$ at $H$ = 0 and the annealing at $T$ for a wait time $t_{w}=3.0\times 10^{4}$ sec. $T$ = 6.5 - 9.0 K. $t$ = 0 is a time at which the field ($H$ = 1 Oe) is applied after the wait time. (b) $T$ dependence of the times $t_{cr1}$ (closed triangles) and $t_{cr2}$ (closed circles) for $t_{w} = 3.0 \times 10^{4}$ sec, where $S_{ZFC}(t)$ has either a peak or kink at $t=t_{cr1}$ and $t_{cr2}$ as shown in Fig.~\ref{fig05}(a)). $H$ = 1 Oe.}
\end{figure}

We have measured the $t$ dependence of $M_{ZFC}$ at the fixed $T$ for various wait time $t_{w}$ [$= (0.2 - 3)\times 10^{4}$ sec], where $H$ = 1 Oe. The measurement was carried out after the ZFC aging protocol: annealing of the system at $T$ = 50 K and $H$ = 0 for $1.2\times 10^{3}$ sec, quenching from 50 K to $T$ at $H$ = 0, and isothermal aging at $T$ for $t_{w}$. The origin of $t$ ($t$ = 0) is a time just after $H$ = 1 Oe is applied at $T$. We find that $M_{ZFC}$ increases with increasing $t$, depending on $t_{w}$ and $T$. Figure \ref{fig05}(a) shows the $t$ dependence of the relaxation rate $S_{ZFC}(t)$ at various $T$, where $t_{w}=3.0\times 10^{4}$ sec. Below $T$ = 7.5 K, $S_{ZFC}(t)$ exhibits a kink at a characteristic time $t_{cr1}\approx (2.06-2.59) \times 10^{3}$ sec and a very broad peak at a characteristic time $t_{cr2}\approx (2.02-2.80)\times 10^{4}$ sec. At $T$ = 8.5 and 9.0 K, $S_{ZFC}(t)$ has two broad peaks at $t=t_{cr1}$ and $t_{cr2}$. Figure \ref{fig05}(b) shows the $T$ dependence of $t_{cr1}$ and $t_{cr2}$ thus obtained. The characteristic time $t_{cr1}$ slightly decreases with increasing $T$. The characteristic time $t_{cr2}$ exhibits a peak around 8 K, and decreases with further increasing $T$. This suggests that the ordered SG phase extending over islands tends to vanish with increasing $T$, in association with the disappearance of the ferromagentic order in each island above $T_{cu}$.

\begin{figure}
\includegraphics[width=7.0cm]{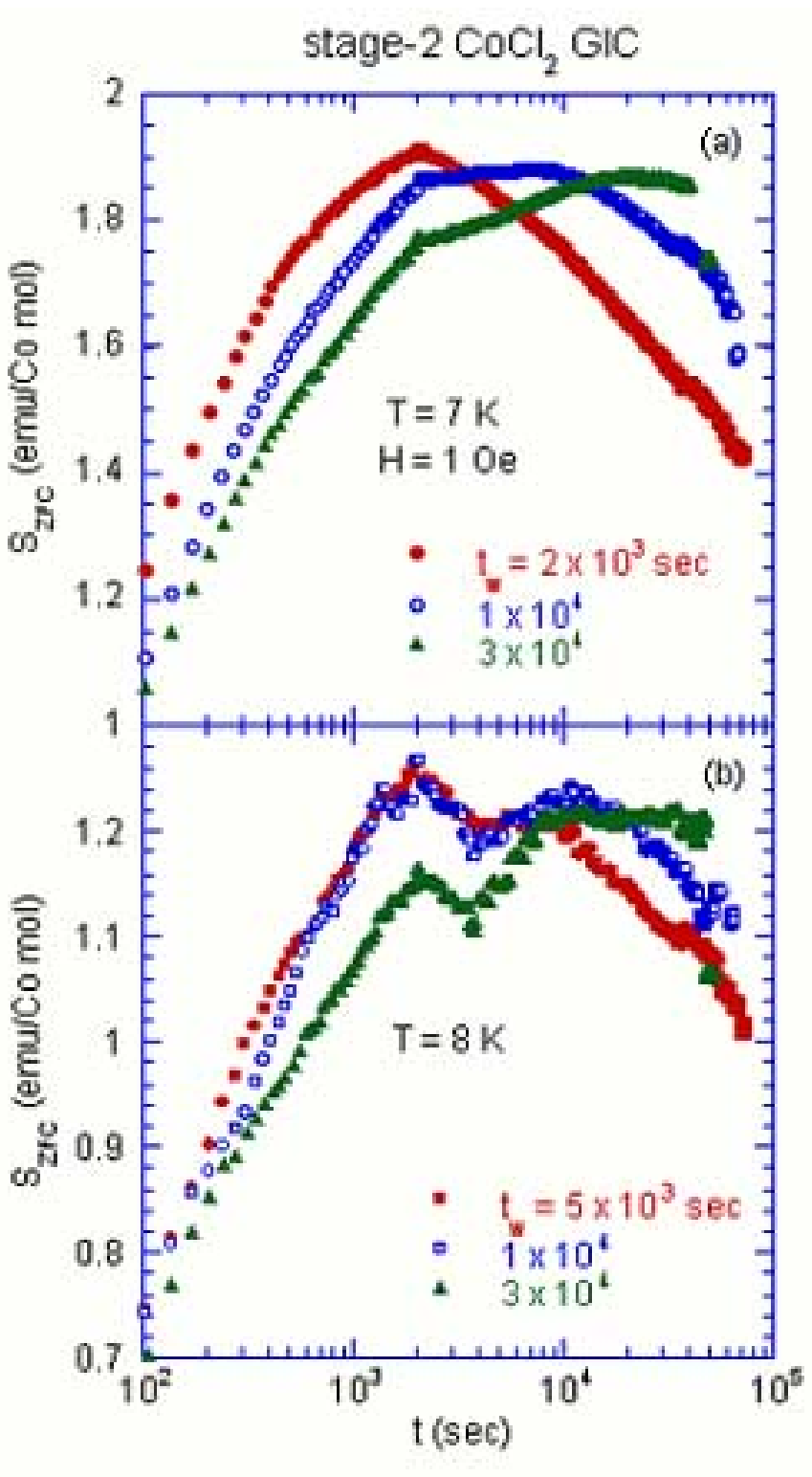}
\caption{\label{fig06}(Color online) $t$ dependence of $S_{ZFC}(t)$ at $T$, where $t_{w}$ is changed as a parameter. (a) $T$ = 7 K and $t_{w}=2.0\times 10^{3}$, $1.0\times 10^{4}$, and $3.0\times 10^{4}$ sec. (b) $T$ = 8 K and $t_{w} = 5.0\times 10^{3}$, $1.0\times 10^{4}$ and $3.0\times 10^{4}$ sec.}
\end{figure}

Figure \ref{fig06}(a) shows the $t$ dependence of $S_{ZFC}(t)$ for different $t_{w}$ at $T$ = 7.0 K, where $H$ = 1 Oe. The $t$ dependence of $S_{ZFC}(t)$ is strongly dependent on $t_{w}$. For $t_{w}=2.0\times 10^{3}$ sec, $S_{ZFC}(t)$ has a peak at $t=t_{cr1}=t_{cr2}=2.1\times 10^{3}$ sec which is very close to $t_{w}$. For $t_{w}=1.0\times 10^{4}$ sec, $S_{ZFC}(t)$ has a kink at $t=t_{cr1}= 2.1\times 10^{3}$ sec and a very broad peak at $t=t_{cr2}=6.9\times 10^{3}$ sec. For $t_{w}= 3\times 10^{4}$ sec, $S_{ZFC}(t)$ has a kink at $t=t_{cr1}=2.1\times 10^{3}$ sec and a very broad peak at $t_{cr2}= 2.04\times 10^{4}$ sec. The time $t_{cr1}$ is independent of $t_{w}$ and nearly equal to $2.1\times 10^{3}$ sec, while the time $t_{cr2}$ is linearly dependent on $t_{w}$: $t_{cr2}/t_{w}=0.68\pm 0.02$. The shift of the peak of $S_{ZFC}(t)$ at $t_{cr2}$ with increasing $t_{w}$ to the long-$t$ side reflects the influence of the aging process on the behavior of the relaxation of the system at $T$ = 7 K, where $t_{cr2}<t_{w}$ Figure \ref{fig06}(b) shows the $t$ dependence of $S_{ZFC}(t)$ for different $t_{w}$ at $T$ = 8.0 K, where $H$ = 1 Oe. The relaxation rate $S_{ZFC}(t)$ exhibits two broad peaks at $t=t_{cr1}$ and $t_{cr2}$ for $t_{w}=5.0\times 10^{3}$, $1.0\times 10^{4}$, and $3.0\times 10^{4}$ sec. The time $t_{cr1}$ is independent of $t_{w}$ and is nearly equal to $2.0\times 10^{3}$ sec, while the time $t_{cr2}$ is linearly dependent on $t_{w}$: $t_{cr2}/t_{w} = 1.123\pm 0.002$. The latter behavior is due to the influence of aging process on the behavior of the relaxation at $T$ = 8 K, where $t_{cr2}>t_{w}$. These aging phenomena suggest that two kinds of ordered domains coexists in the intermediate state. The detail will be discussed in Sec.~\ref{disB}.

\subsection{\label{resultD}Memory effect for $M_{TRM}$ and $M_{ZFC}$}

\begin{figure}
\includegraphics[width=7.0cm]{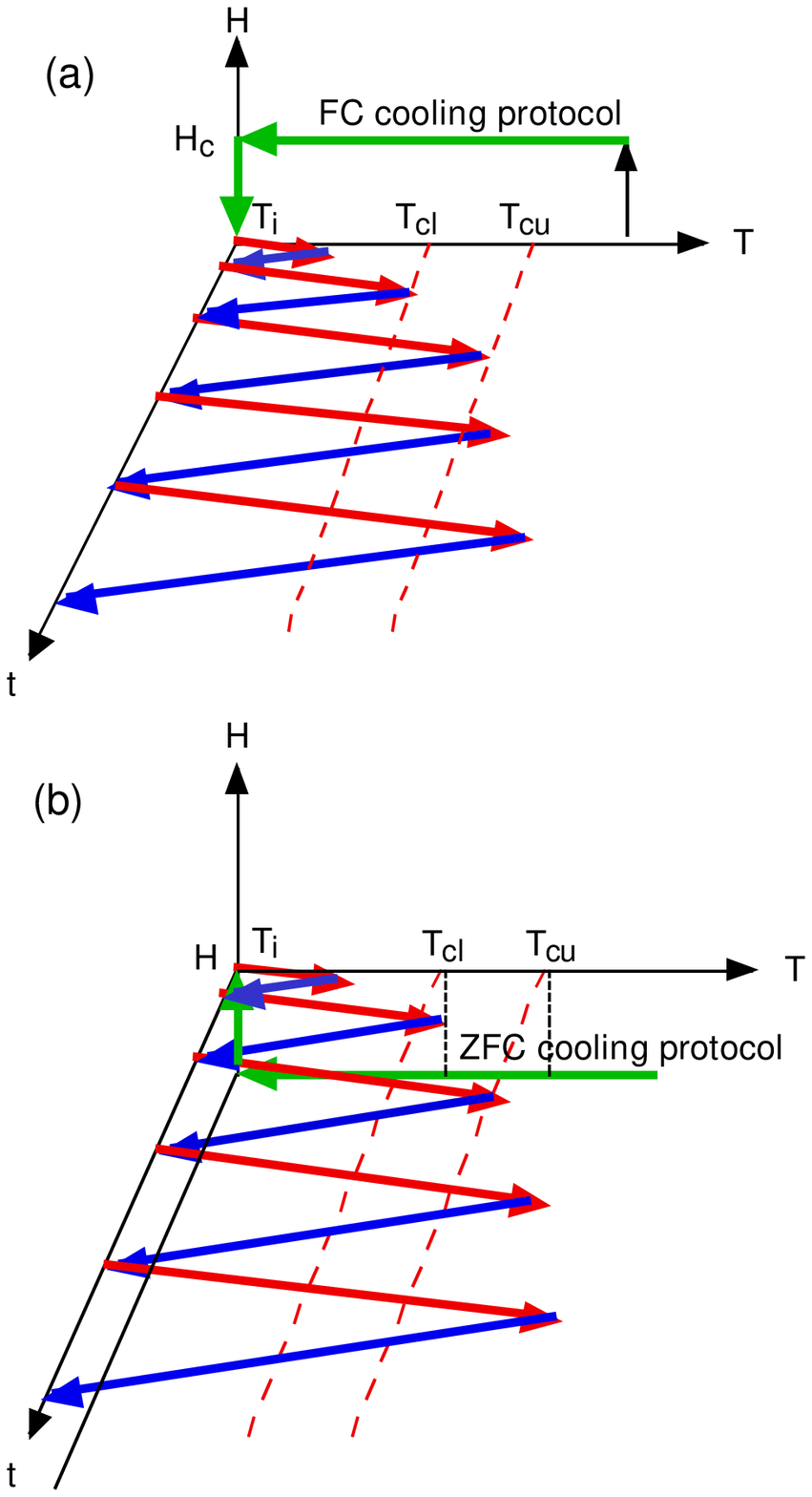}
\caption{\label{fig08}(Color online) (a) TRM case. Repeated processes of heating from $T_{i}$ = 3 K to $T_{r}$ (U-turn temperature), cooling from $T_{r}$ to 3 K, and reheating from $T_{i}$ K to $T_{r+1}$ ($>T_{r}$) in the absence of $H$, after the FC cooling protocol (quenching from 50 K to $T_{i}$ at $H = H_{c} = 1$ Oe and annealing at $T_{i}$ for 100 sec). (b) ZFC case. Repeated processes of heating from $T_{i} = 3$ K to $T_{r}$ (U-turn temperature), cooling from $T_{r}$ to 3 K, and reheating from $T_{i}$ to $T_{r+1}$ ($>T_{r}$) in the presence of $H$ (= 1 Oe), after the ZFC cooling protocol (quenching from 50 K to $T_{i}$ at $H$ = 0 and aging at $T_{i}$ for 100 sec).}
\end{figure}

Here we present our result on memory phenomena of $M_{TRM}$ and $M_{ZFC}$ for stage-2 CoCl$_{2}$ GIC, which is observed in a series of heating and cooling processes. Such a characteristic phenomenon has been predicted theoretically in SG based on a successive bifurcation model of the energy level scheme below the spin freezing temperature.\cite{Matsuura1987b}

(i) \textit{TRM case}.
Before the TRM magnetization measurement, a field cooling (FC) protocol was carried out, consisting of (a) annealing of the system at 50 K for 1200 sec in the presence of $H$ (= 1 or 0.15 Oe), (b) quenching of the system from 50 K to $T = T_{i}$ = 3 K, and (c) aging the system at $T = T_{i}$ at $H$ for a wait time $t_{w}$ = 100 sec. Just after the magnetic field was turned off, the TRM magnetization was measured with increasing $T$ from $T_{i}$ (= 3 K) to $T_{1}$ (= 4.5 K) (the first U-turn temperature) and subsequently with decreasing $T$ from $T_{1}$ to $T_{i}$. (the cooling process). In turn, it was measured with increasing $T$ from $T_{i}$ to $T_{2}$ (= 5.0 K) (the heating process) and subsequently with decreasing $T$ from $T_{2}$ to $T_{i}$ (the cooling process). This process was repeated for the U-turn temperatures $T_{r}$ ($r$ = 3 -11), where $T_{r}>T_{i}$. The schematic diagram of these processes for the TRM measurement is also shown in Fig.~\ref{fig08}(a).

\begin{figure}
\includegraphics[width=7.0cm]{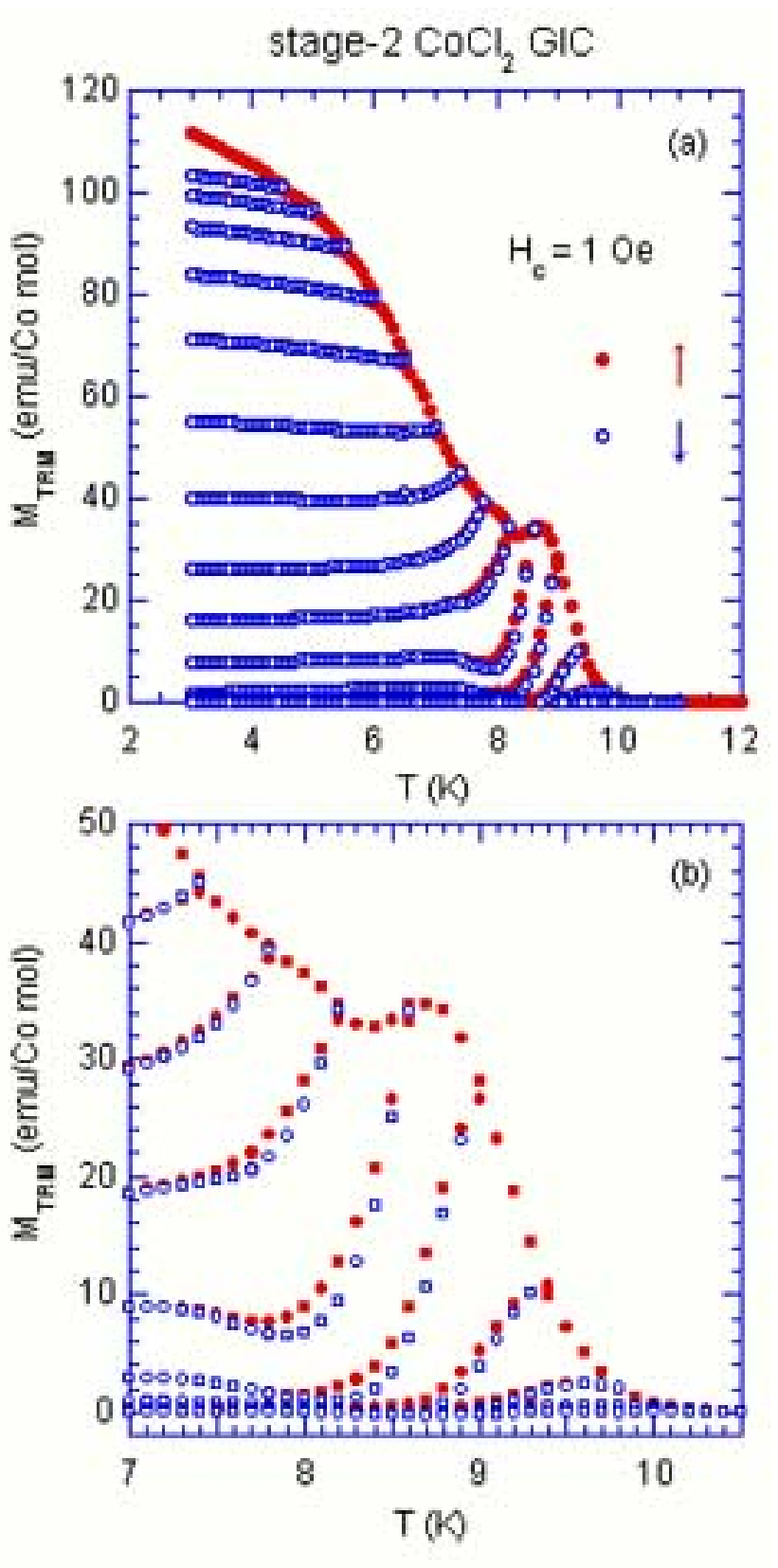}
\caption{\label{fig09}(Color online) (a) and (b) $T$ dependence of $M_{TRM}$ measured at $H$ = 0 in a series of heating (closed circles) and cooling (open circles) process [see Fig.~\ref{fig08}(a) and text for detail] after the FC cooling of the system from 50 to $T_{i}$ = 3 K in the presence of $H_{c}$ (= 1 Oe). Note the data of $M_{TRM}$ shown here is a corrected one by the subtraction of the original data from the $M_{FC}$ part under the remnant magnetic field ($\approx 5$ m Oe).}
\end{figure}

\begin{figure}
\includegraphics[width=7.0cm]{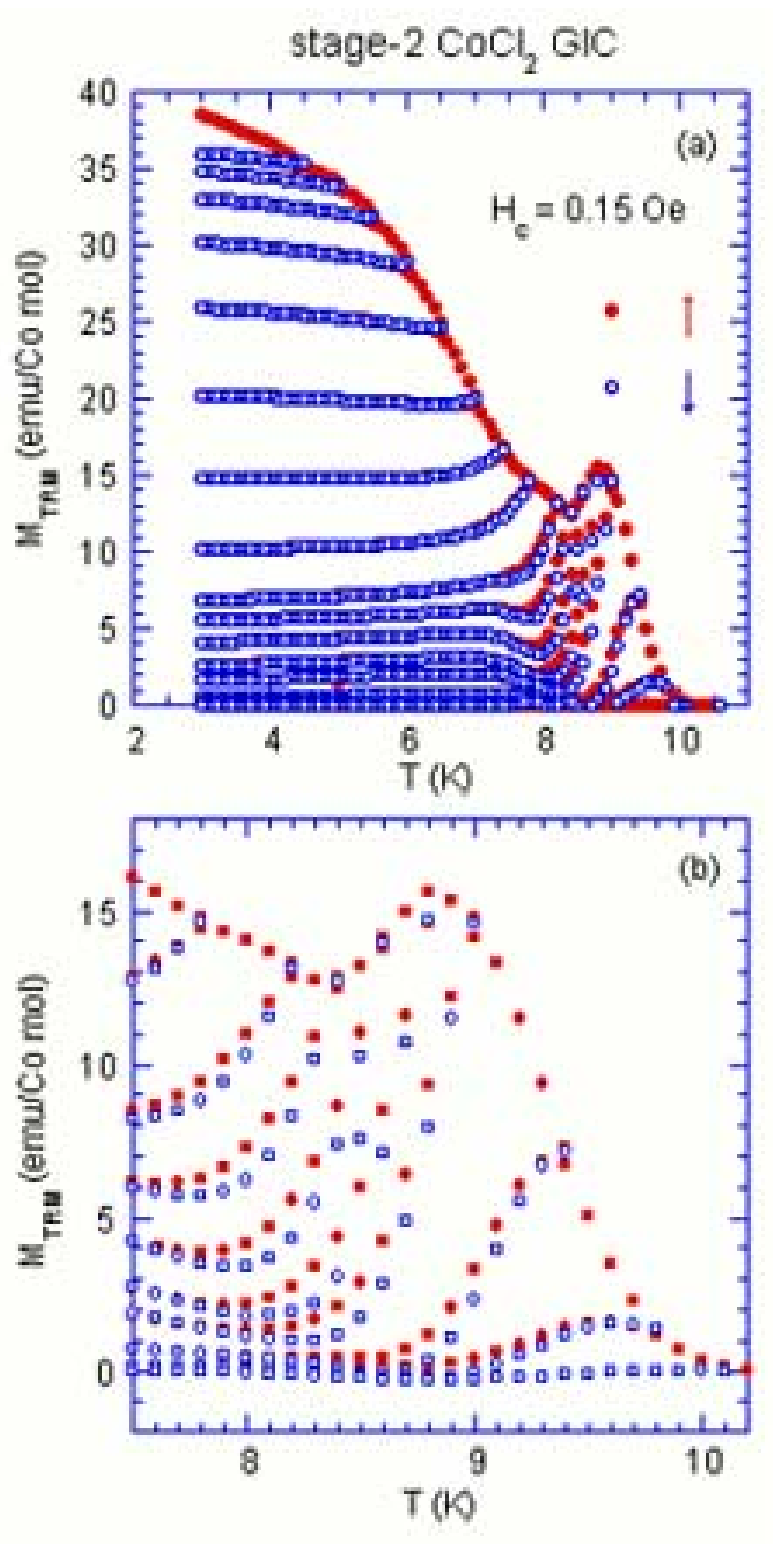}
\caption{\label{fig10}(Color online) (a) and (b) $T$ dependence of $M_{TRM}$ measured at $H$ = 0 in a series of heating (closed circles) and cooling (open circles) processes [see Fig.~\ref{fig08}(a) and the text for detail], after the FC cooling of the system from 50 to $T_{i}$ = 3 K in the presence of $H_{c}$ (= 0.15 Oe). Note the data of $M_{TRM}$ shown here is a corrected one by the subtraction of the original data from the $M_{FC}$ part under the remnant magnetic field ($\approx 5$ m Oe).}
\end{figure}

\begin{figure*}
\includegraphics[width=12.0cm]{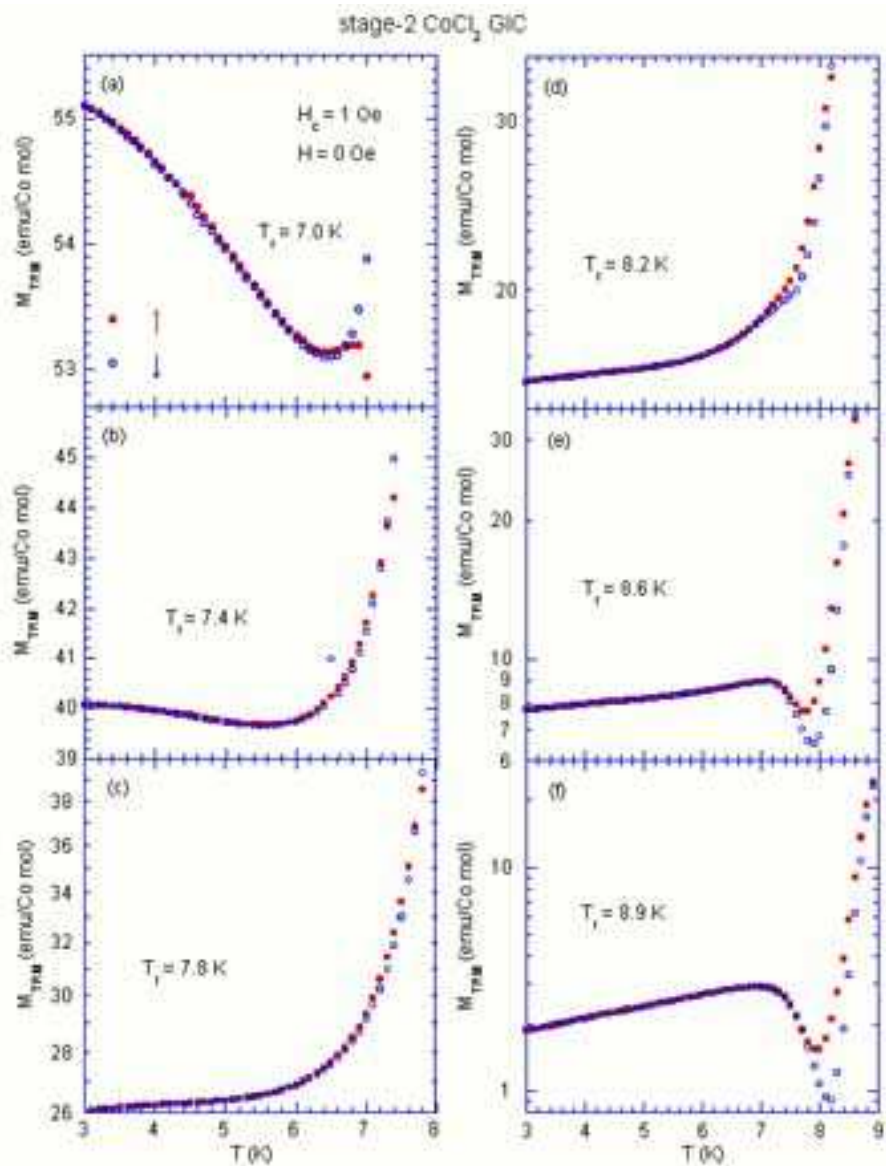}
\caption{\label{fig11}(Color online) $T$ dependence of $M_{TRM}$ at $H$ = 0 which is measured with decreasing $T$ from $T_{r}$ to $T_{i}$ = 3 K [$M_{TRM}(T$$\downarrow)$, open circles] and subsequently measured with increasing $T$ from $T_{i}$ to $T_{r}$ [$M_{TRM}(T$$\uparrow)$, closed circles]. (a) $T_{r}$ = 7.0 K, (b) 7.4 K, (c)7.8 K, (d) 8.2 K, (e) 8.6 K, and (f) 8.9 K.}
\end{figure*}

\begin{figure}
\includegraphics[width=6.5cm]{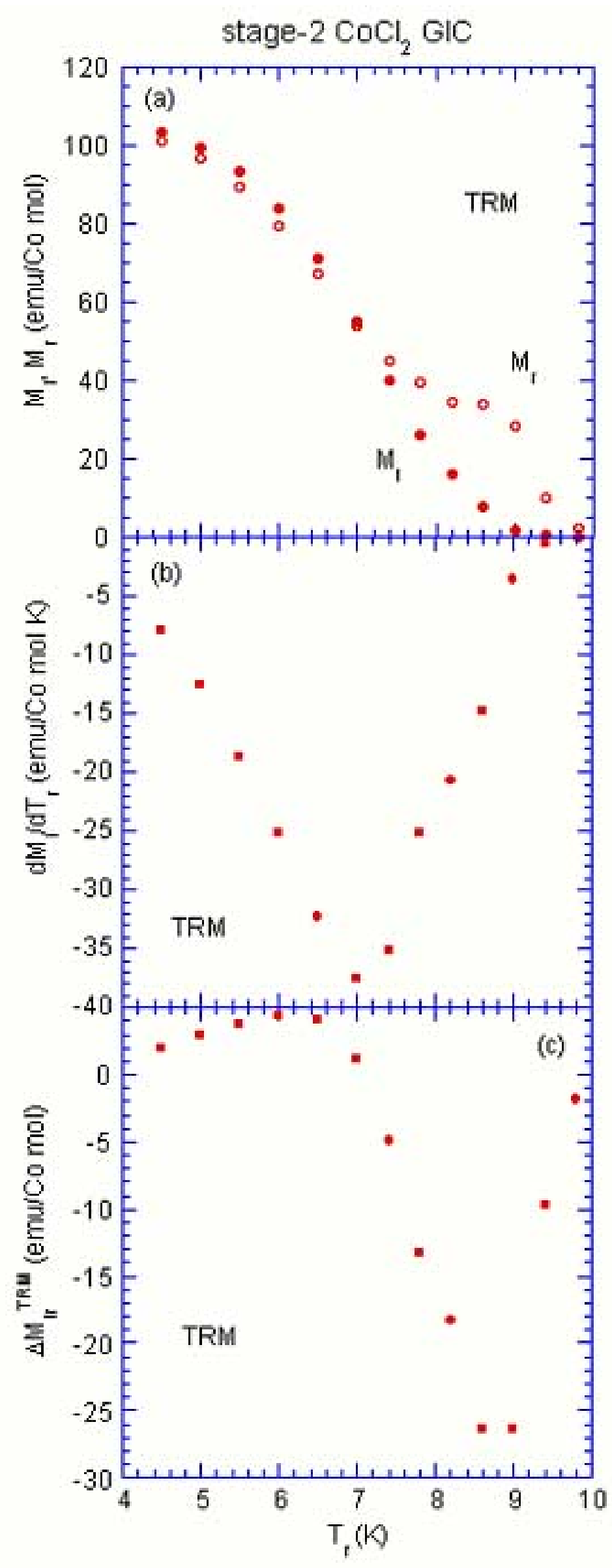}
\caption{\label{fig12}(Color online) (a) $T_{r}$ dependence of $M_{r}$ and $M_{i}$. $M_{r}$ and $M_{i}$ are the value of $M_{TRM}$ at $T=T_{r}$ and $T_{i}$ (= 3 K), respectively, which are obtained in the measurement of $M_{TRM}$ with decreasing $T$ from $T_{r}$ to $T_{i}$ [see Fig.~\ref{fig09}(a)]. (b) T$_{r}$ dependence of d$M_{i}$/d$T_{r}$. (c) $T_{r}$ dependence of $\Delta M_{ir}^{TRM}$ ($=M_{i}-M_{r}$).}
\end{figure}

\begin{figure}
\includegraphics[width=7.0cm]{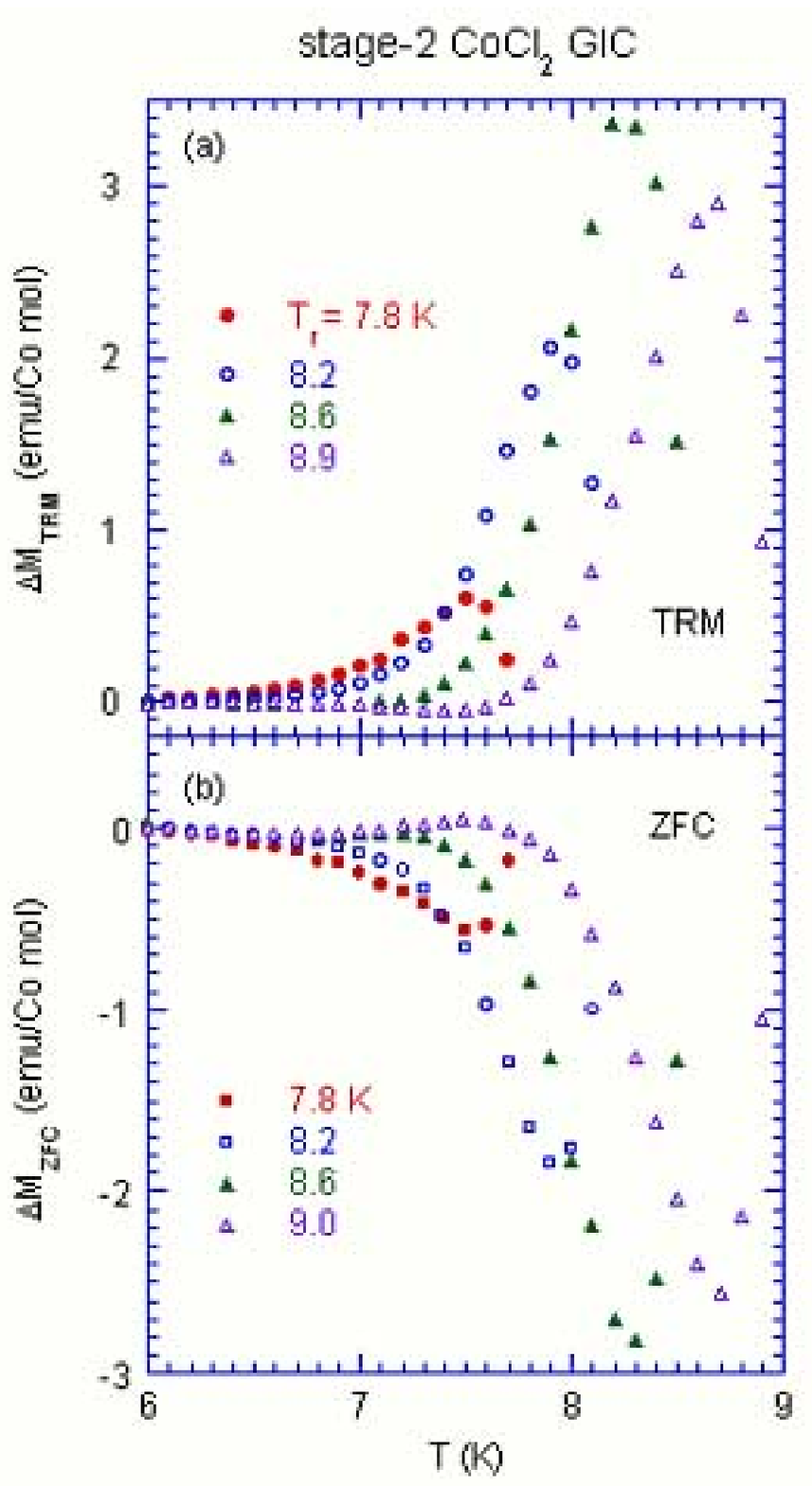}
\caption{\label{fig13}(Color online) (a) $T$ dependence of $\Delta M_{TRM}$ [$=M_{TRM}(T$$\uparrow)-M_{TRM}(T$$\downarrow)$] for 7 K $\leq T \leq T_{r}$. $T_{r}$ = 7.8, 8.2, 8.6, and 8.9 K. (b) $T$ dependence of $\Delta M_{ZFC}$ [$=M_{ZFC}(T$$\uparrow)-M_{ZFC}(T$$\downarrow)$] for 7 K $\leq T \leq T_{r}$. $T_{r}$ = 7.8, 8.2, 8.6, and 9.0 K.}
\end{figure}

Figures \ref{fig09} and \ref{fig10} show typical examples of the $T$ dependence of $M_{TRM}$ obtained using the above procedure, where $H$ = 1 Oe (Figs.~\ref{fig09}(a) and (b)) and $H$ = 0.15 Oe (Figs.~\ref{fig10}(a) and (b)). The value of $M_{TRM}$ lies between those of $M_{TRM}^{ref}$ and $M_{IRM}^{ref}$ (the reference curves) at any $T$ below $T_{cu}$. Here the magnetization $M_{TRM}^{ref}$ is measured with increasing $T$ from $T_{i}$ to 12 K at $H$ = 0 after the FC cooling protocol at $H$ = 1 Oe. The magnetization $M_{IRM}^{ref}$ is measured with increasing $T$ from $T_{i}$ to 12 K at $H$ = 0 after the ZFC cooling protocol from 12 K to $T_{i}$, switching $H$ from 0 to 1 Oe, aging the system at $H$ = 1 Oe for a wait time $t_{w}$ = 100 sec, and again switching $H$ from 1 Oe to 0. Figure \ref{fig11} shows typical examples of the path of $M_{TRM}$ vs $T$ obtained in both the cooling process ($T=T_{r}\rightarrow T_{i}$) and the heating process ($T=T_{i}\rightarrow T_{r}$), where $T_{r}$ = 7.0, 7.4, 7.8, 8.2, 8.6, and 8.9 K. Figure \ref{fig12} (a) shows the $T_{r}$ dependence of $M_{r}$ and $M_{i}$, where $M_{r}$ and $M_{i}$ are the values of $M_{TRM}$ at $T=T_{r}$ and $T_{i}$ in the cooling process ($T=T_{r}\rightarrow T_{i}$), respectively. Figure \ref{fig12}(b) shows the $T_{r}$ dependence of the derivative d$M_{i}$/d$T_{r}$. It exhibits a negative local minimum at $T_{r}=T_{cl}$ = 7 K. Figure \ref{fig12}(c) shows the difference $\Delta M_{ir}^{TRM}$ ($=M_{i}-M_{r}$) which is derived from Fig.~\ref{fig12}(a). For $T_{r}<7.0$ K, the path of $M_{TRM}(T$$\downarrow)$ in the cooling process ($T=T_{r}\rightarrow T_{i}$) is almost the same as that of $M_{TRM}(T$$\uparrow)$ in the heating process ($T=T_{i}\rightarrow T_{r}$), indicating that $M_{TRM}$ vs $T$ curve is reversible on cooling and heating. For $T_{r}\geq 7.8$ K, however, the $M_{TRM}$ vs $T$ curve is irreversible in the cooling and heating processes. For $T_{r}$ = 8.6 and 8.9 K, the curve of $M_{TRM}$ vs $T$ shows a broad peak around $T_{cl}$ and a local minimum between $T_{cl}$ and $T_{cu}$ in both the cooling and heating process between $T_{cl}$ and $T_{cu}$. Figure \ref{fig13}(a) shows the difference $\Delta M_{TRM}$ [$= M_{TRM}(T$$\uparrow)-M_{TRM}(T$$\downarrow)$] as a function of $T$ for various $T_{r}$. The difference $\Delta M_{TRM}$ exhibits a positive peak at a characteristic temperature $T_{p}^{TRM}$ below $T_{r}$ The peak temperature $T_{p}^{TRM}$ linearly increases with increasing $T_{r}$ through the relation $T_{p}^{TRM}-T_{r}=0.3$ K for $7.8\leq T_{r}\leq 8.9$ K.

The intermediate state between $T_{cl}$ and $T_{cu}$ has characteristics of both ordered and disordered states, because of the following reasons. As shown in Fig.~\ref{fig12}(c), $\Delta M_{ir}^{TRM}$ is positive (or $M_{i}>M_{r}$) for $T_{r}<T_{cl}$. It becomes zero when $T_{r}$ is nearly equal to $T_{cl}$, and takes a negative local minimum at $T_{r}=T_{cu}$. The disordered nature of the intermediate state is characterized by the negative value of $\Delta M_{ir}^{TRM}$ (or $M_{i}<M_{r}$). Thus the intermediate state is a sort of disordered state. This feature is in contrast to that of $\Delta M_{ir}^{TRM}>0$ for a ferromagnetic state of normal ferromagnets. However, the intermediate state also exhibits a memory effect which is one of the main features of the ordered state. The value of $M_{TRM}$ at $T_{r}$ almost remains unchanged after the cooling process from $T_{r}$ to $T_{i}$ and the heating process from $T_{i}$ to $T_{r}$. In this sense, the intermediate state is a sort of ordered state: SG ordered phase extending over ferromagnetic islands (see Sec.~\ref{resultA}).

\begin{figure}
\includegraphics[width=7.0cm]{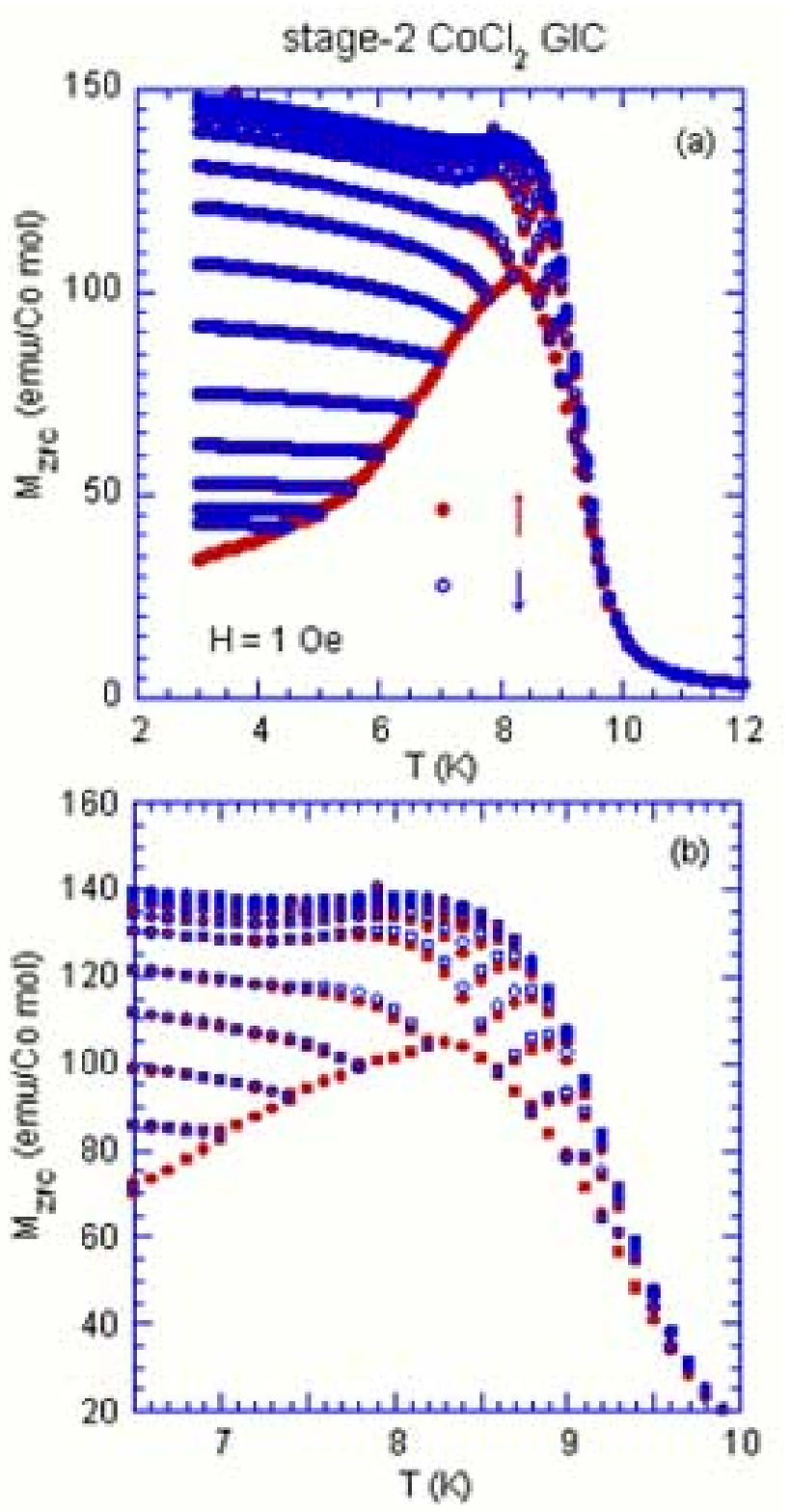}
\caption{\label{fig14}(Color online) (a) and (b) $T$ dependence of $M_{ZFC}$ measured at $H$ = 1 Oe in a series of heating (closed circles) and cooling (open circles) processes [see Fig.~\ref{fig08}(b) and the text for detail], after the ZFC cooling of the system from 50 to 3 K in the absence of $H$.}
\end{figure}

\begin{figure*}
\includegraphics[width=12.0cm]{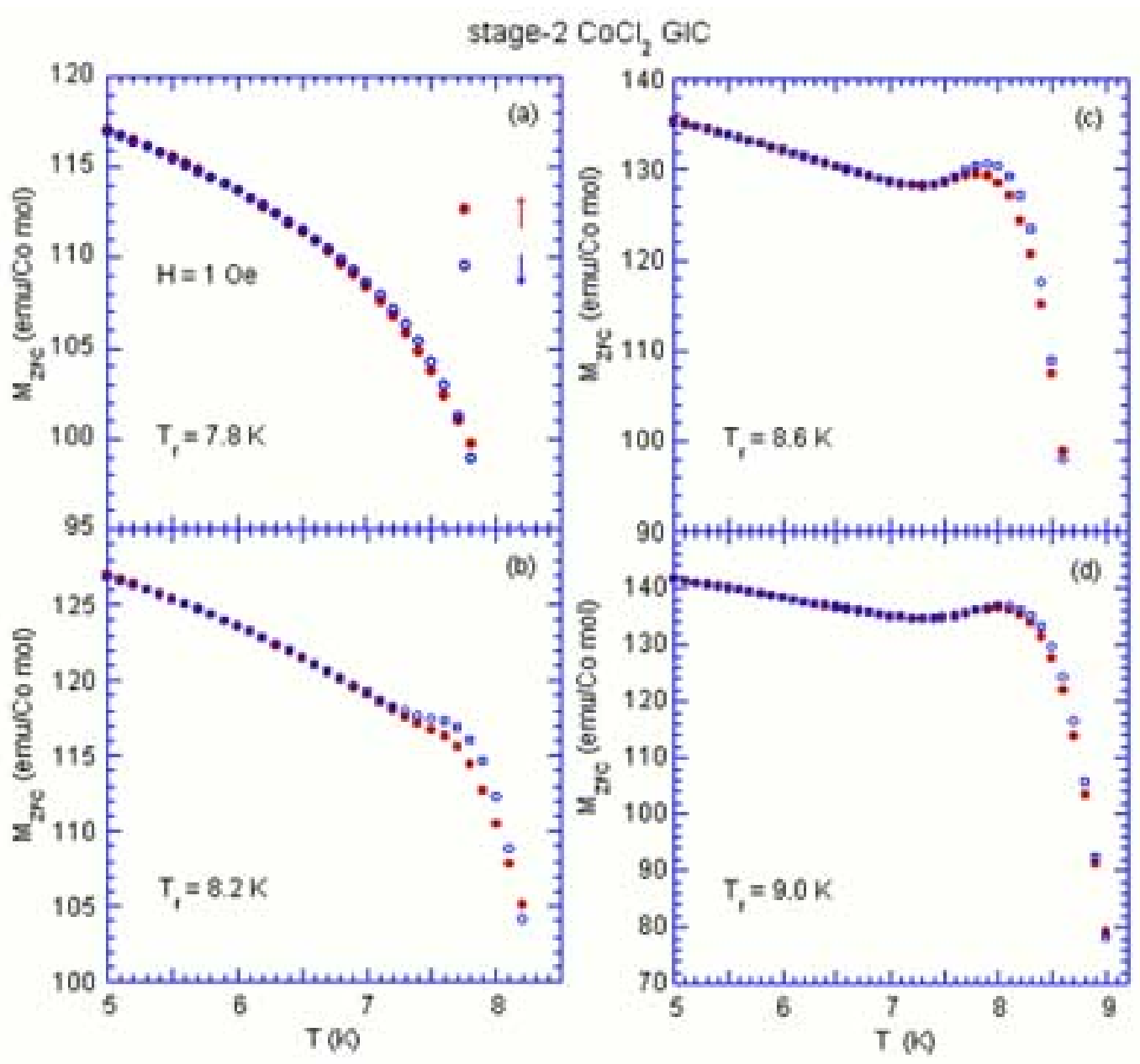}
\caption{\label{fig15}(Color online) $T$ dependence of $M_{ZFC}$ at $H$ = 1 Oe measured with decreasing $T$ from $T_{r}$ to $T_{i}$ (= 3 K) [$M_{ZFC}(T$$\downarrow)$, open circles] and subsequently measured with increasing $T$ from $T_{i}$ to $T_{r}$ [$M_{ZFC}(T$$\uparrow)$, closed circles]. (a) $T_{r}$ = 7.8, (b) 8.2, (c) 8.6, and (d) 8.9 K.}
\end{figure*}

(ii) \textit{ZFC case}. 
Before the ZFC magnetization measurement, a zero-field cooling (ZFC) protocol was carried out. It consists of the following process, (a) annealing of the system at 50 K for 1200 sec in the absence of $H$, (b) quenching of the system from 50 K to $T_{i}$ = 3 K, and (c) aging at $T_{i}$ and $H$ = 0 for a wait time $t_{w}$ = 100 sec. Just after the magnetic field ($H$ = 1 Oe) is applied to the system, the ZFC magnetization $M_{ZFC}$ was measured using the same procedure of heating and cooling: $T_{i}\rightarrow T_{1} \rightarrow T_{i} \rightarrow T_{2} \rightarrow T_{i}\rightarrow T_{3} \rightarrow T_{i} \rightarrow$ and so on, where $T_{r}$ ($r = 1, 2, \cdots$ ) is the U-turn temperature and $T_{r}>T_{i}$. The schematic diagram of these processes is shown in Fig.~\ref{fig08}(b). Figures \ref{fig14}(a) and (b) show the $T$ dependence of $M_{ZFC}$ using the above method. Note that the value of $M_{ZFC}$ lies between those of $M_{ZFC}^{ref}$ and $M_{FC}^{ref}$ at any $T$ below $T_{cu}$. Here $M_{ZFC}^{ref}$ is measured with increasing $T$ from $T_{i}$ to 12 K at $H$ = 1 Oe after the ZFC cooling protocol. The magnetization $M_{FC}^{ref}$ is measured with decreasing $T$ from 12 K to $T_{i}$ in the presence of $H$ (= 1 Oe). Figure \ref{fig15} shows typical examples of the path of $M_{ZFC}$ vs $T$ obtained in both the cooling process ($T=T_{r}\rightarrow T_{i}$) and the heating process ($T=T_{i}\rightarrow T_{r}$), where $T_{r}$ = 7.8, 8.2, 8.6, and 9.0 K. For $T_{r}\leq 7.4$ K, the path of $M_{ZFC}(T$$\downarrow)$ in the cooling process ($T=T_{r}\rightarrow T_{i}$) is the same as that of $M_{ZFC}(T$$\uparrow)$ in the heating process ($T=T_{i}\rightarrow T_{r}$), indicating that $M_{ZFC}$ vs $T$ curve is reversible on cooling and heating. For $T_{r}\geq 7.8$ K, however, the $M_{ZFC}(T)$ curve is irreversible in the cooling and heating processes. For $T_{r}\geq 9.2$ K, both the path of $M_{ZFC}(T$$\downarrow)$ in the cooling process ($T=T_{r}\rightarrow T_{i}$) and the path of $M_{ZFC}(T$$\uparrow)$ in the heating process ($T=T_{i}\rightarrow T_{r}$) coincide with that of $M_{FC}^{ref}$ which is obtained by cooling from the PM phase to $T=T_{i}$ in the presence of $H$ (= 1 Oe).

\begin{figure}
\includegraphics[width=6.5cm]{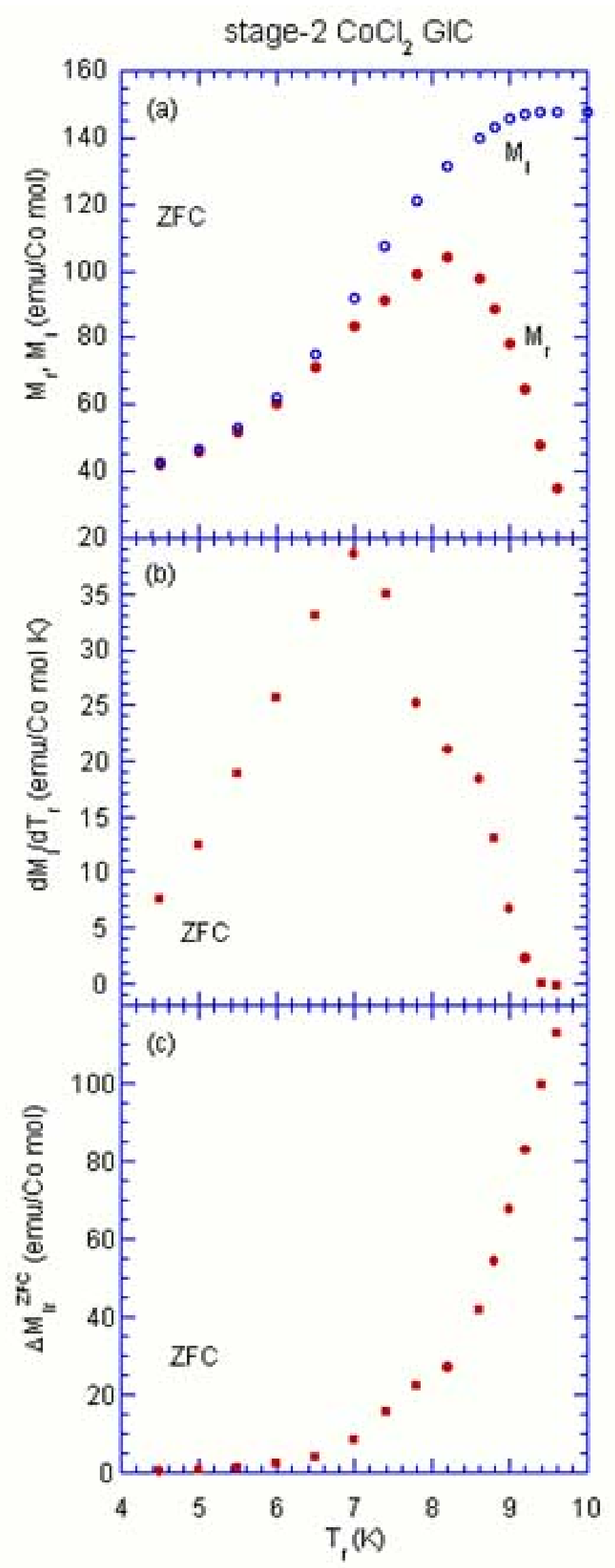}
\caption{\label{fig16}(Color online) (a) $T_{r}$ dependence of $M_{r}$ and $M_{i}$. $M_{r}$ and $M_{i}$ are the value of $M_{ZFC}$ at $T=T_{r}$ and $T_{i}$ (= 3 K), respectively, which are obtained in the measurement of $M_{ZFC}$ with decreasing $T$ from $T_{r}$ to $T_{i}$ [see Fig.~\ref{fig13}(a)]. (b) $T_{r}$ dependence of d$M_{i}$/d$T_{r}$. (c) $T_{r}$ dependence of $\Delta M_{ir}^{ZFC}$ ($= M_{i}-M_{r}$).}
\end{figure}

Figure \ref{fig16} (a) shows the $T_{r}$ dependence of $M_{r}$ and $M_{i}$, where $M_{r}$ and $M_{i}$ are the values of $M_{ZFC}$ at $T=T_{r}$ and $T_{i}$ in the cooling process ($T=T_{r}\rightarrow T_{i}$), respectively. Figure \ref{fig16}(b) shows the $T_{r}$ dependence of the derivative d$M_{i}$/d$T_{r}$. It exhibits a local maximum at $T_{r}=T_{cl}$ = 7 K. Figure \ref{fig16}(c) shows the difference $\Delta M_{ir}^{ZFC}$ ($= M_{i}-M_{r}$) which is derived from Fig.~\ref{fig16}(a). For $T_{r}<T_{cl}$, $\Delta M_{ir}^{ZFC}$ is positive but is nearly equal to zero: $M_{i}\approx M_{r}$. At $T_{r}=T_{cl}$, $\Delta M_{ir}^{ZFC}$ starts to increase with increasing $T_{r}$. At $T_{r}\approx 8.2$ K where $M_{r}$ exhibits a peak, $\Delta M_{ir}^{ZFC}$ drastically increases with further increasing $T_{r}$. For $T_{r}>T_{cu}$, $M_{i}$ reaches the equilibrium value. Note that $M_{i}$ is 56\% value at $T_{r}=T_{cl}$ and 98\% at $T_{r}=T_{cu}$ of the equilibrium value. This result also supports that the intermediate state has the nature of the ordered state.

Figure \ref{fig13}(b) shows the difference $\Delta M_{ZFC}$ [$= M_{ZFC}(T$$\uparrow)-M_{ZFC}(T$$\downarrow)$] as a function of $T$ for various $T_{r}$. The difference $\Delta M_{ZFC}$ exhibits a negative local minimum at a characteristic temperature $T_{p}^{ZFC}$ below $T_{r}$. The $T$ dependence of $\Delta M_{ZFC}$ is almost the same that of $-\Delta M_{TRM}$. Note that $T_{p}^{ZFC}$ for the ZFC process is exactly the same as $T_{p}^{TRM}$ for the TRM process at the same $T_{r}$. In other words, the magnetization gained in the TRM process corresponds to the magnetization lost in the ZFC process. This result supports the fundamental relation 
\begin{equation} 
M_{ZFC}(t_{w},t) = M_{FC}(0,t+t_{w})-M_{TRM}(t_{w},t),
\label{eq03} 
\end{equation} 
between the time relaxation of TRM, FC, and ZFC magnetization at low $H$ in SG's.\cite{Djurberg1995} Similar behavior is observed in the genuine TRM and ZFC measurements of spin glass Ag(11 at \% Mn).\cite{Mathieu2001a}

\subsection{\label{resultE}Memory effect for $M_{FC}$}

\begin{figure}
\includegraphics[width=7.0cm]{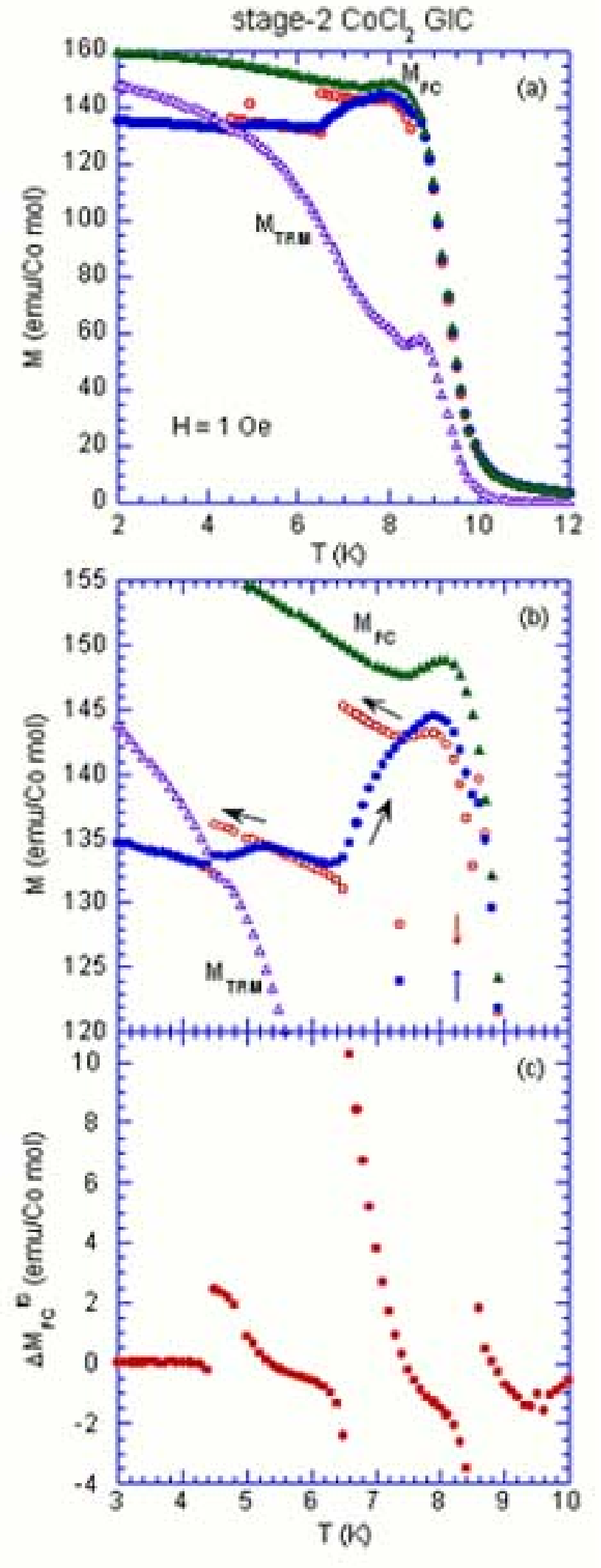}
\caption{\label{fig17}(Color online) (a) and (b) $T$ dependence of $M_{FC}^{IS}(T$$\downarrow)$ and $M_{FC}^{IS}(T$$\uparrow)$ observed in the following FC cooling protocol. The system was quenched from 50 to 12 K in the presence of $H$ (= 1 Oe). $M_{FC}^{IS}(T$$\downarrow)$ was measured with decreasing $T$ from 12 to 1.9 K but with intermittent stops at $T$ = 8.5, 6.5, and 4.5 K for a wait time $t_{w}=3.0\times 10^{4}$ sec. The field is cut off during each stop. $M_{FC}^{IS}(T$$\uparrow)$ was measured at $H$ = 1 Oe with increasing $T$ after the above cooling process. The $T$ dependence of $M_{FC}^{ref}$ and $M_{TRM}^{ref}$ are also shown as reference curves. These are measured after the FC cooling protocol without intermittent stop (reference curves). (c) $T$ dependence of the difference $\Delta M_{FC}^{IS}$ [$= M_{FC}^{IS}(T$$\downarrow)-M_{FC}^{IS}(T$$\uparrow)$].}
\end{figure}

We present a peculiar memory effect observed in our system using a unique FC cooling protocol. Similar behavior is observed in spin glasses and superparamagnets.\cite{Sun2003,Sasaki2005} The result is shown in Fig.~\ref{fig17}. First our system was cooled through the FC cooling protocol from 50 K to intermittent stop temperatures $T_{s}$ (= 8.5, 6.5, 4.5 K) in the presence of $H_{c}$ = 1 Oe. When the system was cooled down to each $T_{s}$, the field was cut off ($H$ = 0) and aged at $T_{s}$ for $t_{w}$ ($=3.0\times 10^{4}$ sec). In this case, the magnetization $M_{FC}^{IS}(T$$\downarrow)$ decreases with time due to the relaxation. After the wait time $t_{w}$ at $T_{s}$, the field ($H_{c}$ = 1 Oe) was applied again and the FC cooling process was resumed. Such a FC cooling process leads to a step-like behavior of $M_{FC}^{IS}(T$$\downarrow)$ curve. The value of $M_{FC}^{IS}(T$$\downarrow)$ after resuming below the lowest stop temperature behaves almost in parallel to that of the FC magnetization without the intermittent stops (as reference curve). After reaching 1.9 K, the magnetization $M_{FC}^{IS}(T$$\uparrow)$ was measured in the presence of $H$ (= 1 Oe) as the temperature is increased at the constant rate (0.05 K/min). The magnetization $M_{FC}^{IS}(T$$\downarrow)$ thus measured exhibits a broad peak at a characteristic temperature $T_{a}$ = 5.3 K and a peak at $T_{a}$ = 7.9 K. The spin configuration imprinted at the intermittent stop at $T_{s}$ for a wait time $t_{w}$ at $H$ = 0 during the FC cooling process strongly affects the $T$ dependence of $M_{FC}^{IS}(T$$\uparrow)$ when the temperature is increased, exhibiting a peculiar memory effect. Figure \ref{fig17}(c) shows the $T$ dependence of the difference $\Delta M_{FC}^{IS}$ between $M_{FC}^{IS}(T$$\uparrow)$ and $M_{FC}^{IS}(T$$\downarrow)$. Such an oscillatory behavior in $\Delta M_{FC}^{IS}$ has been reported in superparamagnets, superspin glasses, and spin glasses. Sasaki et al.\cite{Sasaki2005} have shown that the aging and memory effects originate solely from a broad distribution of relaxation times. This model may be true for the SG phase extending over ferromagnetic islands. When the field is cut off at $T=T_{s}$ for $t_{s}$ ($=3\times 10^{4}$ sec), the magnetic moments of the system formed of small islands whose relaxation times are longer than $t_{s}$ are frozen in when the cooling is restarted. These frozen states are reactivated when the system is reheated at $T_{s}$.

\section{\label{dis}DISCUSSION}
\subsection{\label{disA}Ordered and disordered nature of intermediate state}
From our result on the memory effect for $M_{TRM}$ presented in Sec.~\ref{resultD}, we find that the intermediate state between $T_{cl}$ and $T_{cu}$ has characteristics of both ordered and disordered states. The disordered nature of the intermediate state is characterized by the negative value of $\Delta M_{ir}^{TRM}$ (or $M_{i}<M_{r}$). This feature is in contrast to that of $\Delta M_{ir}^{TRM}>0$ for a ferromagnetic state of normal ferromagnets. However, the intermediate state also exhibits a memory effect as an ordered state. The value of $M_{TRM}$ at $T_{r}$ almost remains unchanged after the cooling process from $T_{r}$ to $T_{i}$ and the heating process from $T_{i}$ to $T_{r}$. In this sense, the intermediate is a sort of ordered state. Such a characteristic phenomenon is explained in terms of intermediate state with intraisland-order and partial interisland-order. But if the intermediate state is a completely interisland-disorder state, then thermal equilibrium $M_{TRM}$ in the state should be always zero. The memory effects observed here indicates that the intermediate state is a SG phase extending over islands, where each island is ferromagnetically ordered. The observed characteristic two-step ordering could be understood thermodynamically as a hierarchy one that the SG phase extending over islands occurs at $T_{cu}$ and the 3D order occurs at $T_{cl}$ as an equilibrium state, due to the enhanced interplanar interaction. 

\subsection{\label{disB}The domain size from the relaxation rate $S_{ZFC}(t)$}
The aging behavior of the intermediate state after the ZFC aging protocol can be understood in terms of the droplet model.\cite{Fisher1986,Bray1987,Fisher1988} This ZFC aging protocol process to the intermediate state is completed at $t_{a}$ = 0, where $t_{a}$ is defined as an age (the total time after the ZFC aging protocol process). Then the system is aged at $T$ under $H$ = 0 until $t_{a}=t_{w}$, where $t_{w}$ is a wait time. Correspondingly, the size of domain defined by $R_{T}(t_{a})$ grows with the age of $t_{a}$ and reaches $R_{T}(t_{w})$ just before the field is turned on at $t$ = 0 or $t_{a}=t_{w}$. The aging behavior in $M_{ZFC}$ is observed as a function of the observation time $t$. After $t$ = 0, a probing length $L_{T}(t)$ corresponding to the maximum size of excitation grows with $t$, in a similar way as $R_{T}(t_{a})$. When $L_{T}(t)\ll R_{T}(t_{w})$, quasi-equilibrium dynamics is probed, but when $L_{T}(t)\gg R_{T}(t_{w})$, non-equilibrium dynamics is probed. It is theoretically predicted that the mean domain-size $L_{T}(T)$ is described by a power law given by 
\begin{equation} 
L_{T}(t)/L_{0} \approx (t/t_{0})^{1/z(T)},
\label{eq04} 
\end{equation} 
where $L_{0}$ and $t_{0}$ are microscopic length and time scale and the $T$-dependent exponent $z(T)$. It is predicied that the relaxation rate $S_{ZFC}(t)$ exhibits a peak when $L_{T}(t)\approx R_{T}(t_{w})$.\cite{Lundgren1990}

In Sec.~\ref{resultC} we show that $S_{ZFC}(t)$ exhibits two peak at $t=t_{cr1}$ and at $t=t_{cr2}$ in the intermediate state. Here the time $t_{cr1}$ is much shorter than $t_{cr2}$. The characteristic time $t_{cr1}$ is independent of the wait time $t_{w}$, while $t_{cr2}$ is linearly dependent on $t_{w}$. The existence of $t_{cr1}$ and $t_{cr2}$ suggests that there are two kinds of characteristic domains with sizes $L_{T}(t=t_{cr1})$ and $L_{T}(t=t_{cr2})$. The domain size $L_{T}(t=t_{cr1})$ may correspond to the size of small island. The growth of the in-plane correlation length $\xi_{a}$ is limited by the size of the small island. In contrast, $t_{cr2}$ tends to increase with increasing $t_{w}$. The domain size $L_{T}(t=t_{cr2})$ is much larger than that of the small island. These domains may be formed of several small islands through the RKKY interactions. This implies that $\xi_{a}$ is actually much larger than the size of small islands.

\subsection{\label{disC}In-plane spin correlation length from the genuine TRM measurement}
Using the fundamental link given by Eq.(\ref{eq04}), the time dependence of the relaxation rate $S_{TMR}(t)$ for the TRM magnetization can be obtained from $S_{TRM}(t)$ as $S_{TRM}(t)=-S_{ZFC}(t)$. From the result of Figs.~\ref{fig06}(a) and (b), it follows that that $S_{TRM}(t)$ with t$_{w}= 1.0\times 10^{4}$ sec exhibits two negative local minima at $t_{cr1}$ ($= 2.1\times 10^{3}$ sec) and $t_{cr2}$ ($= 6.9\times 10^{3}$ sec) at 7 K and $t_{cr1}$ ($= 2.0\times 10^{3}$ sec) and $t_{cr2}$ ($= 1.12\times 10^{4}$ sec) at 8 K. In the genuine TRM measurement (see Sec.~\ref{resultB}), the system is aged at $T=T_{s}$ for $t=t_{s}= 1.0\times 10^{4}$ sec in the presence of $H=H_{c}$ (= 1 Oe). The above result of $S_{TRM}(t)$ with $t_{w}=1.0\times 10^{4}$ sec indicates that two kinds of domains coexist: domain with small island size and large domains formed of small islands through the inter-island interactions. Since $t_{cr2}$ is on the same order as $t_{s}$ ($= t_{w}$) or a little shorter than $t_{s}$, the size of ordered domains reaches the in plane spin correlation length $\xi_{a}(T_{s})$ in thermal equilibrium. The ordered domains are frozen in when the cooling is resumed. These frozen states are reactivated when the system is reheated at $T_{s}$. The magnetization $\Delta M_{TRM}(T=T_{s};T_{s},t_{s})$ is approximated by $g_{a}\mu_{B}S[\xi_{a}(T_{s})/a)]^{2}$, where $g_{a}$ (= 6.4)  is the Land\'{e} $g$-factor of Co$^{2+}$ spin along the $a$ axis in the $c$ plane, $S$ (= 1/2) is a fictitious spin, and $a$ is the in-plane lattice constant.\cite{Enoki2003} As shown in Fig.~\ref{fig04}(a), $(\Delta M_{TRM})_{max}$ increases with decreasing $T_{s}$ from $T_{cu}$ to $T_{cl}$. Correspondingly $\xi_{a}(T_{s})$ increases with increasing $T_{s}$ and tends to diverge at $T=T_{cl}$.

\subsection{\label{disD}Overlap length from the genuine TRM measurement}
Here we discuss the effect of the overlap length on the genuine TRM magnetization (Sec.~\ref{resultB}). The ordered domains generated at $T=T_{s}$ are frozen in and survives the spin reconfiguration occurring at lower temperature on shorter length scales. The rejuvenation of the system occurs as the temperature is decreased away from $T_{s}$. The spin configuration imprinted at $T_{s}$ is recovered on reheating. In this sense, the system sustains a memory of an equilibrium state reached after a stop-wait process at $T_{s}$. The influence of the spin configuration imprinted at a stop-wait protocol is limited to a restricted temperature range around $T_{s}$ on reheating. The width of this region may be assigned to the existence of an overlap between the spin configuration attained at $T_{s}$ and the corresponding state at a very neighboring temperature ($T_{s}+\Delta T$).

Here we introduce a concept of the overlap which is encountered in the SG system.\cite{Fisher1986,Bray1987,Fisher1988} The SG equilibrium configurations at different temperatures at $T$ and $T+\Delta T$ are strongly correlated only up to the overlap length $L_{\Delta T}$, beyond which these correlations decay to zero. From the droplet theory, the overlap length $L_{\Delta T}$ is described by
\begin{equation} 
L_{\Delta T}/L_{0} \approx (T^{1/2}\vert \Delta T\vert/\Upsilon_{T}^{3/2})^{-1/\zeta },
\label{eq05} 
\end{equation} 
where $\zeta$ is the chaos exponent ($\zeta =d_{s}/2-\theta$), $d_{s}$ is the fractal dimension of the surface of the droplet, and $\Upsilon_{T}$ is the temperature corresponding to the wall stiffness $\Upsilon$. The overlap length decreases with increasing $\vert\Delta T\vert$. The width $\Delta T$ is determined from the condition that $R_{T_{s}}(t_{s})=L_{\Delta T}$.

In our system, the spin configuration imprinted during the stop-wait protocol at $T=T_{s}$ for $t=t_{s}$ is unaffected by a small temperature shift such that the overlap length $L_{\Delta T}$ is larger than the average domain sizes. There is a sufficient overlap between the equilibrium spin configurations at the two temperatures $T_{s}$ and $T_{s}+\Delta T$. The situation is different when the temperature shift becomes large. The overlap length becomes shorter than the original domain sizes. A smaller overlap between spin configurations promotes the formation of broken domains. When the temperature shift is sufficiently large, the overlap length is much shorter than the original domain sizes, leading to the rejuvenation of the system.\cite{Lundgren1990,Granberg1988,Jonsson2002,Suzuki2004} The asymmetric form of $\Delta M_{TRM}(T;T_{s},t_{s})$ around $T=T_{s}$ (= 7.6 K) between $T_{cl}$ and $T_{cu}$ indicates that the spin configuration under the positive $T$-shift is different from that under the negative $T$-shift. The domain size of the partial SG phase may drastically increase with decreasing $T$ below $T=7.6$ K, partly because of enhanced interisland interactions.

\section{CONCLUSION}
We have studied the aging dynamics of stage-2 CoCl$_{2}$ GIC between two magnetic phase transitions at $T_{cl}$ (= 7.0 K) and $T_{cu}$ (= 8.9 K). The observed aging phenomena is well explained within the framework of the droplet model for SG systems. The intermediate state between $T_{cl}$ and $T_{cu}$ has characteristic of both ordered and disordered states. A genuine thermoremnant magnetization (TRM) measurement indicates that the memory of the specific spin configurations imprinted at temperatures between $T_{cl}$ and $T_{cu}$ during the field-cooled (FC) aging protocol can be recalled when the system is re-heated at a constant heating rate. The zero-field cooled (ZFC) and TRM magnetization is examined in a series of heating and reheating process. The magnetization shows both characteristic memory and rejuvenation effects. The time $(t)$ dependence of the relaxation rate $S_{ZFC}(t)$ after the ZFC aging protocol with a wait time $t_{w}$, exhibits two peaks at characteristic times $t_{cr1}$ and $t_{cr2}$ between $T_{cl}$ and $T_{cu}$. An aging process is revealed as the strong $t_{w}$ dependence of $t_{cr2}$. These results suggest that two types of ordered domains coexist in the intermediate state. The intermediate state is a SG phase extending over ferromagentic islands. Below $T_{cl}$, the interplanar interaction between islands in adjacent intercalate layers becomes strong, leading to the 3D ordered phase. 

\begin{acknowledgments}
The authors would like to thank H. Suematsu for providing them with single crystal of kish graphite.
\end{acknowledgments}

\end{document}